\documentclass[a4paper,11pt,oneside,fleqn]{article}

\usepackage[mathscr]{eucal}
\usepackage{amsmath,amssymb,amsthm}
\usepackage{graphicx,subfig}
\usepackage{caption}
\usepackage{color}

\usepackage{cite}
\usepackage{hyperref}
\hypersetup{colorlinks, linkcolor=blue, citecolor=blue, urlcolor=blue}

\allowdisplaybreaks

\makeatletter
\long\def\@makecaption#1#2{%
  \vskip\abovecaptionskip\footnotesize
  \sbox\@tempboxa{#1. #2}%
  \ifdim \wd\@tempboxa >\hsize
    #1. #2\par
  \else
    \global \@minipagefalse
    \hb@xt@\hsize{\hfil\box\@tempboxa\hfil}%
  \fi
  \vskip\belowcaptionskip}
\makeatother

\newcommand{\bfPhi}{\boldsymbol{\Phi}}
\newcommand{\bfxi}{\boldsymbol{\xi}}

\newcommand{\bfn}{\mathbf{n}}

\newcommand{\bfx}{\mathbf{x}}
\newcommand{\unitn}{\hat{\mathbf{n}}}
\newcommand{\bfnabla}{\boldsymbol{\nabla}}

\newcommand{\bfB}{\mathbf{B}}
\newcommand{\bfD}{\mathbf{D}}
\newcommand{\bfF}{\mathbf{F}}

\newcommand{\bfK}{\mathbf{K}}
\newcommand{\bfN}{\mathbf{N}}

\newcommand{\rmd}{\mathrm{d}}

\title{A spectral-infinite-element solution of Poisson's equation: an application to self gravity}
\author{Hom Nath Gharti$^{1,}$\thanks{Email: hgharti@princeton.edu} and Jeroen Tromp$^{1,2}$\\
${}^1$ Department of Geosciences, Princeton University,\\
Princeton, NJ 08544, USA\\
${}^2$ Program in Applied \& Computational Mathematics, Princeton University,\\
Princeton, New Jersey, USA}

\begin{document}

\maketitle

\begin{abstract}
We solve Poisson's equation by combining a spectral-element method with a mapped infinite-element method.
We focus on problems in geostatics and geodynamics, where Earth's gravitational field is determined by Poisson's equation inside the Earth and Laplace's equation in the rest of space.
Spectral elements are used to capture the internal field, and infinite elements are used to represent the external field.
To solve the weak form of Poisson/Laplace equation, we use Gauss-Legendre-Lobatto quadrature in spectral elements inside the domain of interest. Outside the domain, we use Gauss-Radau quadrature in the infinite direction, and Gauss-Legendre-Lobatto quadrature in the other directions.
We illustrate the efficiency and accuracy of the method by comparing the gravitational fields of a homogeneous sphere and the Preliminary Reference Earth Model (PREM) with
(semi-)analytical solutions.
\end{abstract}


\newpage

\section{Introduction}
During post-seismic and post-glacial relaxation, mantle convection, tidal-loading, and long-period seismic wave propagation,
Earth's motions displace mass, thereby inducing dynamic gravitational changes.
Addressing such problems requires joint solution of the conservation laws of continuum mechanics
and Poisson's equation.
Outside the Earth, where the mass density vanishes, gravity is governed by Laplace's equation.
Because gravity decays to zero at infinity,
the essential boundary condition is defined at infinity,
and we are forced to numerically solve an unbounded differential equation.

In simulations of global seismic wave propagation,
gravitational perturbations induced by particle motions are ignored, and only the unperturbed equilibrium gravitational field is taken into account~\cite{komatitsch2002a, komatitsch2002b}.
This co-called ``Cowling approximation'' is justified for relatively short-periods waves (periods less than $\sim$250~s), but is invalid for global quasistatic problems and free-oscillation seismology~\cite{dahlen1998}. For example, in simulations of glacio-isostatic adjustments, gravity plays a central role, and one must solve Poisson's equation in conjunction with conservation of mass and momentum.
Few approaches are available to solve such problems for general three-dimensional (3D) Earth models.

One simple approach is to consider a very large domain that includes portions of outer space.
This strategy requires large computational resources and is often inaccurate~\cite{tsynkov1998}. A higher order solver based on the convolution integral was also proposed to solve unbounded Poisson equation \cite{hejlesen2013}.
Alternatively, one may use spherical harmonics to represent the solution to Poisson/Laplace's equation.
Such an approach is used in geodynamics and coupled-mode simulations of seismic wave propagation~\cite{dahlen1974,wu1982,tromp1999a,tromp1999b}.
Finally, spherical harmonics may be used to capture the gravitational field outside the Earth, while the internal field is solved using finite/spectral elements.
Such a mixed representation has been used in post-glacial rebound calculations~\cite{peltier1978,zhong2003,geruo2013,alattar2014} and for simulations of global seismic wave propagation~\cite{chaljub2004}, requiring fast Legendre transformations.
Finite elements and spherical harmonics do not naturally couple, and numerical implementations often require significant iterations to reach convergence~\cite{latychev2005}.

In solid and fluid mechanics, the displacement descent approach~\cite{bettess1977,medina1983,elesnawy1995} and the coordinates ascent approach~\cite{beer1981,zienkiewicz1983,kumar1985,angelov1991} are both widely used to solve unbounded problems.
In the displacement descent approach an element in the physical domain is mapped to an element in a natural domain of interval $[0, \infty]$.
This is achieved by multiplying the standard interpolation functions by suitable decay functions.
Since the integration interval is $[0, \infty]$, classical Gauss-Legendre quadrature cannot be employed.
Either Gauss-Legendre quadrature has to be modified to accommodate the $[0, \infty]$ interval, or Gauss-Laguerre quadrature can be used~\cite{mavriplis1989}. In terms of programming, both the Jacobian of the mapping and the numerical quadrature have to be modified from the classical finite-element method. 
The coordinate ascent approach is also referred to as the ``mapped infinite element'' approach.
In this technique an element that extends to infinity in the physical domain is mapped to a standard natural element with interval $[-1, 1]$.
This is achieved by defining shape functions using a point outside the element opposite to the infinite direction; this point is often called a pole.
The corresponding shape function possesses a singularity at the far end of the infinite element to map the infinite location.
Standard Gauss-Legendre quadrature can be used.
In terms of programming, only the Jacobian of the mapping needs to be modified, and quadrature identical to the classical finite-element method can be used.

The spectral-element method (SEM) is a higher-order finite element method which uses nodal quadrature, specifically, Gauss-Legendre-Lobatto quadrature.
For dynamical problems, the mass matrix is diagonal by construction.
The method is highly accurate and efficient,
and is widely used in applications involving wave propagation~\cite{faccioli1997,peter2011,tromp2008,seriani2008}, fluid dynamics ~\cite{patera1984,canuto1988,deville2002}, as well as for quasistatic problems~\cite{gharti2012a,gharti2012b}.

To solve the unbounded Poisson/Laplace equation efficiently,
we combine the infinite element approach based on coordinate ascent with the spectral-element method.
From hereon, we refer to this method as a ``spectral-infinite-element method''.
This paper details the development and implementation of the spectral-infinite-element method,
and is illustrated and validated based on calculations of the gravitational fields of a homogeneous sphere and the Preliminary Reference Earth Model (PREM).

\section{Formulation}

\subsection{Governing equation}

\begin{figure}[htbp]
\centering
\includegraphics[scale=0.5]{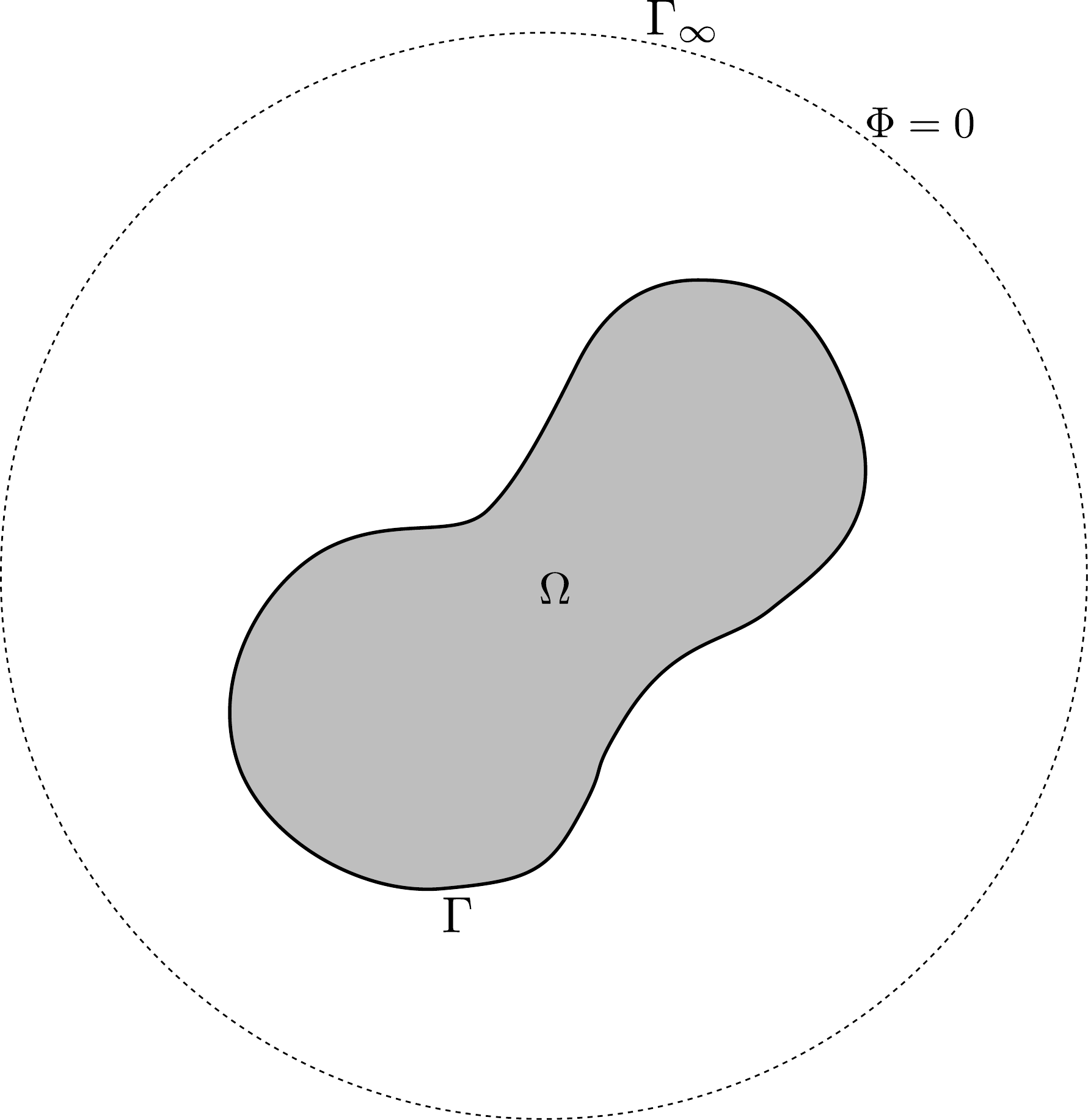}
\caption{Schematic diagram of an unbounded domain.}
\label{fig:domain}
\end{figure}

We denote the domain of interest by~$\Omega$, and its boundary by $\Gamma$ (see Figure~\ref{fig:domain}).
All of space is denoted by~$\bigcirc$, and the ``boundary'' at infinity is denoted by~$\Gamma_\infty$.
The gravitational field, $\Phi$, is governed by Poisson's equation \cite{dahlen1998},
\begin{equation}
\nabla^2\Phi=4\,\pi\,G\,\rho  \quad\text{in}\quad \bigcirc 
\quad,
\label{eq:gov}
\end{equation}
where $G$ denotes the universal gravitational constant and $\rho$ the mass density.
Outside the domain~$\Omega$ the mass density vanishes, and the governing equation~(\ref{eq:gov}) reduces to Laplace's equation: $\nabla^2\Phi=0$.
The associated boundary conditions are
\begin{equation}
\begin{split}
[\Phi]_-^+ &= 0 \quad\text{on}\quad \Gamma \quad ,\\
[\hat{\bfn}\cdot\bfnabla\Phi]_-^+ &= 0 \quad\text{on}\quad \Gamma \quad ,
\end{split}
\label{eq:bc}
\end{equation}
where~$[\,\cdot\,]_-^+$ denotes the jump in the enclosed quantity when going from the $-$~side of boundary~$\Gamma$ to the $+$~side,
and where~$\unitn$ denotes the unit outward normal to the boundary, pointing from the $-$~side to the $+$~side.
If the model domain~$\Omega$ has internal discontinuities
---as is the case at phase boundaries in the Earth's interior---
the gravitational field satisfies  boundary conditions~(\ref{eq:bc}) across each discontinuity.
At $\Gamma_\infty$ the field satisfies the boundary condition
\begin{equation}
\Phi = 0 \quad\text{on}\quad \Gamma_\infty \quad .
\label{eq:bc2}
\end{equation}

\subsection{Discretization}

The weak form of the governing equation~(\ref{eq:gov}) subject to boundary conditions~(\ref{eq:bc}) and~(\ref{eq:bc2}) is
\begin{equation}
\int_\bigcirc\bfnabla w\cdot\bfnabla\Phi\,\rmd V = \mbox{}-4\,\pi\,G\int_\Omega w\,\rho\,\rmd V+\int_{\Gamma_\infty} w\,\hat{\bfn}\cdot\bfnabla\Phi\,\rmd S\quad,
\label{eq:weak}
\end{equation}
where $w$ denotes a test function, and where we have used the fact that the mass density vanishes outside of~$\Omega$.

\begin{figure}[htbp]
\centering
\includegraphics[scale=0.5]{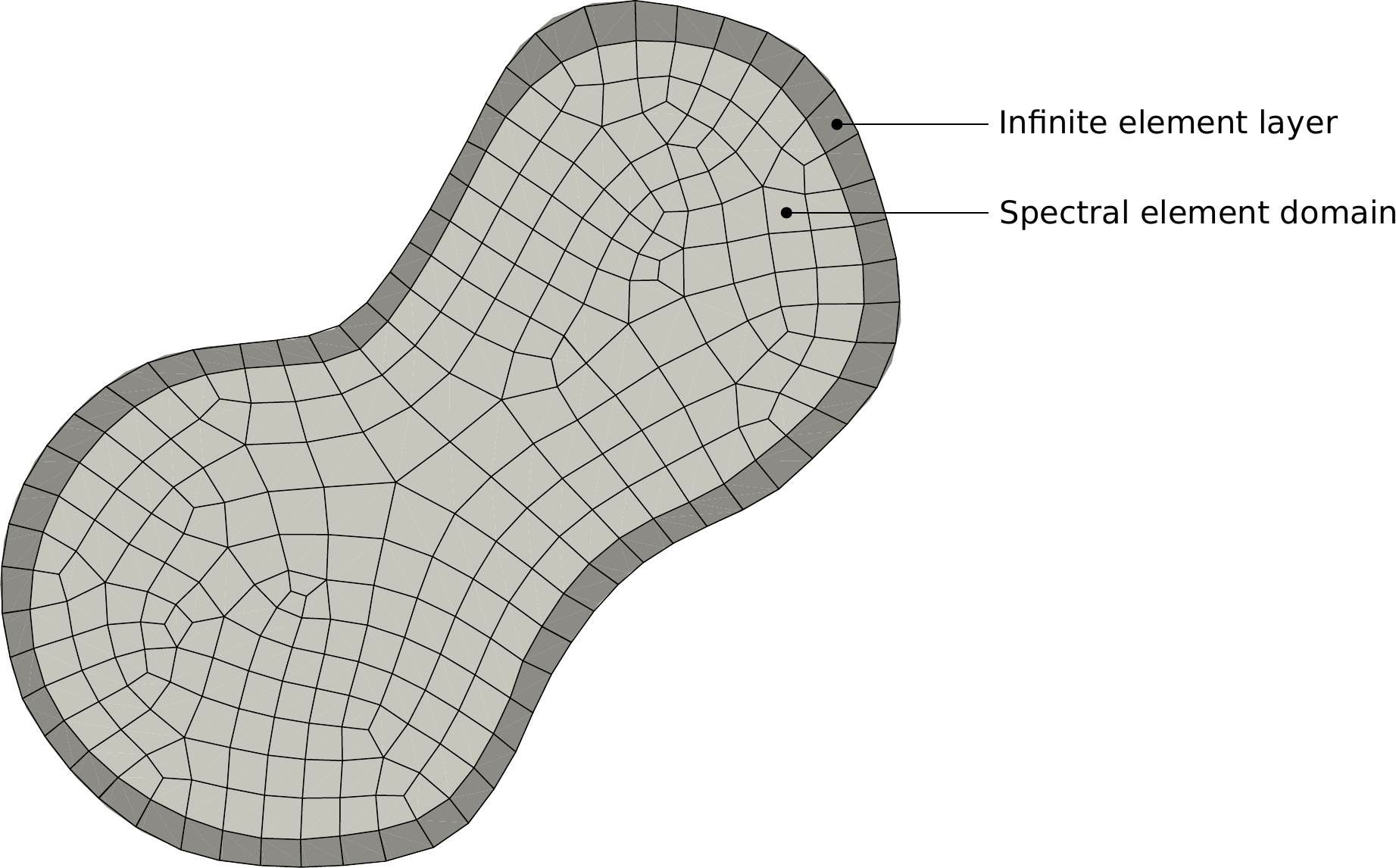}
\caption{The domain of interest, $\Omega$, is discretized using spectral elements (light grey elements). A single layer of infinite elements is added outside the domain (dark grey elements).}
\label{fig:inflayer}
\end{figure}

The domain~$\Omega$ is meshed using spectral elements (Figure~\ref{fig:inflayer}).
A single layer of infinite elements is added outside the domain in order to reproduce the behavior of outer space.
As we discuss in more detail later,
spectral and infinite elements share the same interpolation functions, namely Lagrange polynomials, but use different quadrature.
Thus, the gravitational potential field~$\Phi$ is discretized in natural coordinates~$\bfxi$ as
\begin{equation}
\Phi(\bfxi)=\sum_{\alpha=1}^n \Phi_\alpha\,N_\alpha(\bfxi)
\quad,
\label{eq:interp}
\end{equation}
where, $\Phi_\alpha$ denotes the field value at quadrature point~$\bfxi_\alpha$ and~$N_\alpha$ an interpolation function.
The total number of quadrature points in an element is denoted by~$n$, and is given by the product of the number of quadrature points in each dimension, $n^j$, $j=1,2,3$, that is, $n=\prod_{j=1}^3 n^j$.
The interpolation functions $N_\alpha$ in natural coordinates are determined by the tensor product of one-dimensional Lagrange polynomials, that is
\begin{equation}
N^j_{\alpha^j}(\xi^j)=\prod_{\substack{\beta=1 \\ \beta\ne\alpha^j}}^{n^j} \frac{(\xi^j-\xi^j_\beta)}{(\xi^j_{\alpha^j}-\xi^j_\beta)}
\quad ,
\label{eq:lag1d}
\end{equation}
such that
\begin{equation}
N_\alpha(\bfxi)=\prod_{j=1}^3\,N^j_{\alpha^j}(\xi^j)
\quad .
\label{eq:lag}
\end{equation}
Here $\alpha$ denotes the index of quadrature point~$\bfxi_\alpha=\{\xi_{\alpha^1},\xi_{\alpha^2},\xi_{\alpha^3}\}$.

The test function~$w$ is taken to be an interpolation function~$N_\alpha$, making the approach a Galerkin method.
Upon substituting $w=N_\alpha$ and $\Phi$ given by equation~(\ref{eq:interp}), in equation~(\ref{eq:weak}), we obtain a set of elemental linear
equations that may be written conveniently in matrix-vector form:
\begin{equation}
\bfK_e\,\bfPhi_e=\bfF_e
\quad.
\end{equation}
The quantities $\bfK_e$ and $\bfF_e$ are known, respectively, as the stiffness matrix and force
vector of an element. Similarly, $\bfPhi_e$ is the gravitational potential vector. Symbolically, we write
\begin{equation}
\begin{split}
\bfK_e &= \int_{\Omega_e} \bfB^T\,\bfB\,\rmd V \quad ,\\
\bfF_e &= 4\,\pi\,G\int_{\Omega_e} \rho\,\bfN_e\,\rmd V+\int_{\Gamma_\infty} (\hat{\bfn}\cdot\bfnabla\Phi)\,\bfN_e\,\rmd S
\quad,
\end{split}
\end{equation}
where
\begin{equation}
\bfB=\bfD\,\bfN_e^T \quad ,
\end{equation}
such that
\begin{equation}
\bfD=\left\lbrace\frac{\partial}{\partial x} \quad \frac{\partial}{\partial y} \quad \frac{\partial}{\partial z}\right\rbrace^T
\quad,
\end{equation}
and
\begin{equation}
\bfN_e=\{ N_1 \quad N_2 \quad N_3 \cdots N_n \}^T
\quad,
\end{equation}
\begin{equation}
\bfPhi_e=\{\Phi_1 \quad \Phi_2 \quad \Phi_3 \cdots \Phi_n\}^T
\quad.
\end{equation}

After assembling the elemental matrices and vectors, we obtain a set of global linear equations
\begin{equation}
\bfK\,\bfPhi=\bfF
\quad,
\end{equation}
where $\bfK$ and $\bfF$ are known, respectively, as the global stiffness matrix and global force
vector. Similarly, $\bfPhi$ is the global gravitational potential vector.

On the boundary~$\Gamma_\infty$, gravity vanishes, but the surface integral~$\int_{\Gamma_\infty} w\,\hat{\bfn}\cdot\bfnabla\Phi\,\rmd S$ is unknown,
because $\bfnabla\Phi$ is unknown;
we call the corresponding unknown part of the global force vector~$\bfF_1$, and its known part, determined by $\mbox{}-4\,\pi\,G\int_\Omega w\,\rho\,\rmd V$, $\hat{\bfF}_2$.
Similarly, the global potential vector~$\bfPhi$ is split in terms of a known component, $\hat{\bfPhi}_1$, which corresponds to values on~$\Gamma_\infty$ which are zero, and an unknown component, $\bfPhi_2$, which corresponds to the rest of the domain.
Therefore we may partition the global equation as follows:
\begin{equation}
\begin{bmatrix}
\bfK_{11} & \bfK_{12} \\
\bfK_{21} & \bfK_{22}
\end{bmatrix} \begin{bmatrix}
\hat{\bfPhi}_1 \\
\bfPhi_2
\end{bmatrix} = \begin{bmatrix}
\bfF_1 \\
\hat{\bfF}_2
\end{bmatrix}
\quad,
\end{equation}
resulting in two sets of linear equations, namely,
\begin{equation}
\bfK_{11}\, \hat{\bfPhi}_1 + \bfK_{12}\, \bfPhi_2 = \bfF_1\quad.
\label{eq:part1}
\end{equation}
and
\begin{equation}
\bfK_{21}\, \hat{\bfPhi}_1 + \bfK_{22}\, \bfPhi_2 = \hat{\bfF}_2\quad,
\label{eq:part2}
\end{equation}

First, the unknown part $\bfPhi_2$ is obtained by rewriting equation~(\ref{eq:part2}) as
\begin{equation}
\bfK_{22}\, \bfPhi_2 = \hat{\bfF}_2 - \bfK_{21}\, \hat{\bfPhi}_1 \quad,
\end{equation}
leading to a system of linear equations for~$\bfPhi_2$.
Once the $\bfPhi_2$ is determined, $\bfF_1$ can be directly obtained from equation~(\ref{eq:part1}).

\subsection{Mapping}

\begin{figure}[ht]
\centering
\subfloat[]{\includegraphics[scale=0.32]{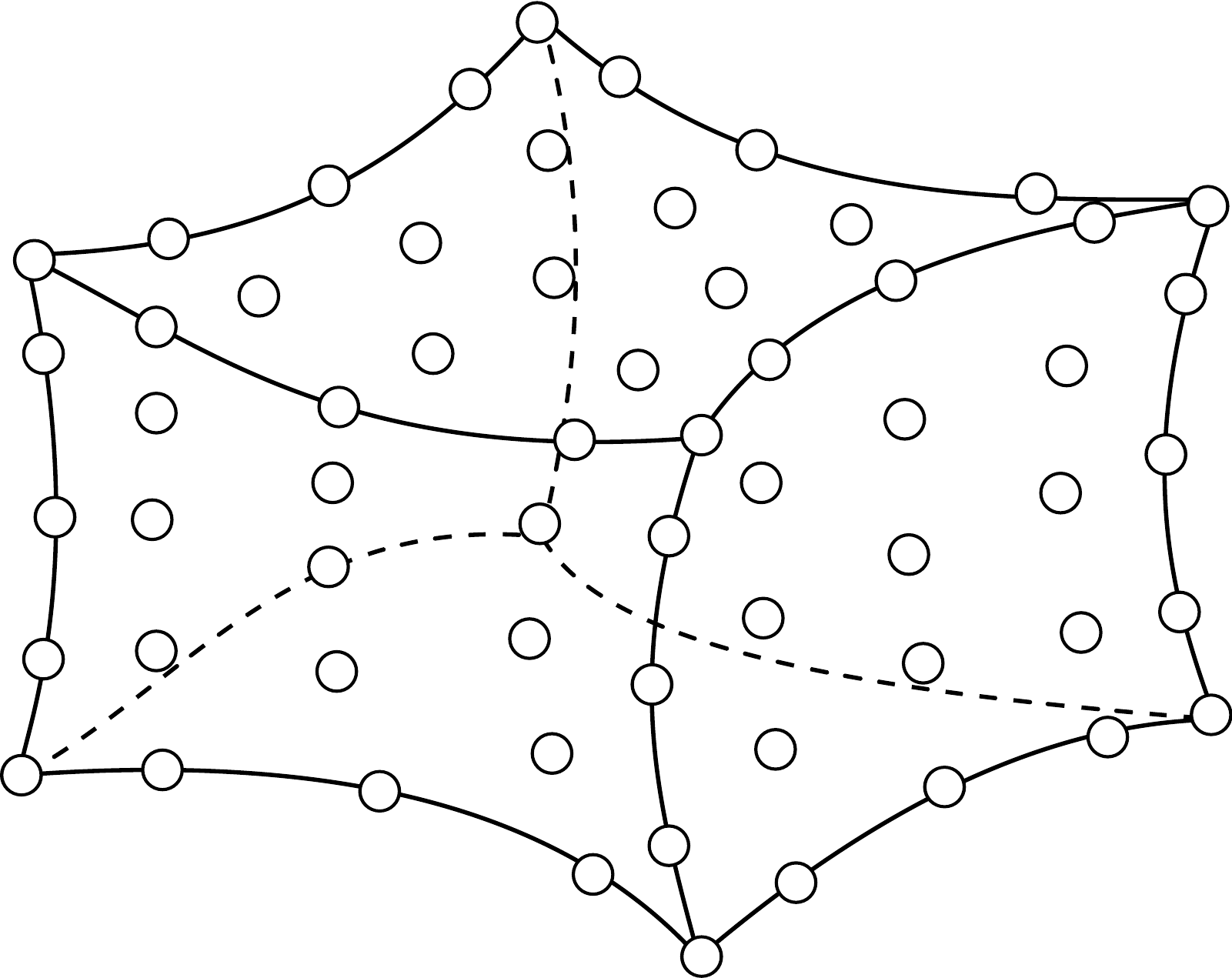}} \hspace{2.0em}
\subfloat[]{\includegraphics[scale=0.32]{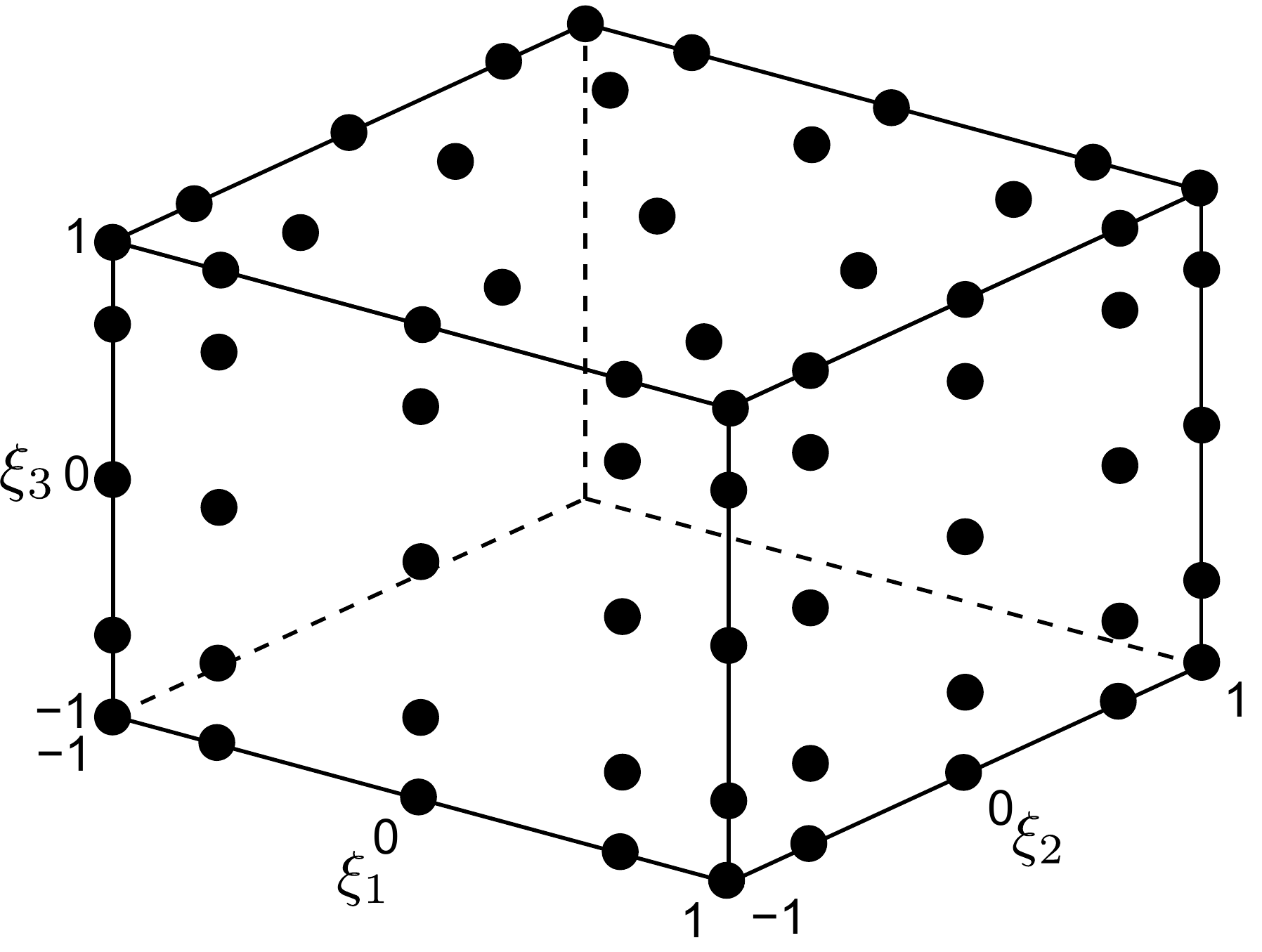}}
\caption{(a) A typical spectral element with five interpolation nodes in each dimension (open circles).
(b) The same spectral element mapped to its natural coordinates.
Gauss-Legendre-Lobatto points (solid black circles) are used for numerical integration.
Integration points and interpolation nodes coincide. Only interpolation nodes on the three visible faces are shown for clarity.}
\label{fig:spectral_element}
\end{figure}

For numerical integration, a point $\bfx=\lbrace x^i\rbrace$ in a physical element
is mapped to a point $\bfxi=\lbrace \xi^j\rbrace$ in the natural element, as illustrated in
Figures~\ref{fig:spectral_element}a-b.
This mapping is different for the spectral-element and infinite-element domains, as described below.

\subsubsection*{Spectral elements}

A spectral element is mapped to a natural element using the transformation
\begin{equation}
\bfx(\bfxi)=\sum_{\alpha=1}^{n_\text{g}} \bfx_\alpha\,M_\alpha(\bfxi)
\quad .
\label{eq:interp_geom}
\end{equation}
Here $M_\alpha$ denotes a shape function and $n_\text{g}$ the number of geometrical nodes, $\bfx_\alpha$, of an element.
The shape function $M_\alpha$ is defined similar to the interpolation function (Equation~\ref{eq:interp}).
However, the number of interpolation points, $n$, and the number of geometrical points, $n_g$, may differ.
In general, $n_g<n$ for the spectral-element method, leading to a sub-parametric formulation.
The Jacobian matrix of the transformation is determined using the relation
$J^{ij}(\bfxi)=\partial x^i(\bfxi)/\partial \xi^j$.
We use Gauss-Legendre-Lobatto (GLL) quadrature for numerical integration, in which the interpolation and quadrature points are identical.

\subsubsection*{Infinite elements}

\begin{figure}[ht]
\centering
\subfloat[]{\includegraphics[scale=0.28]{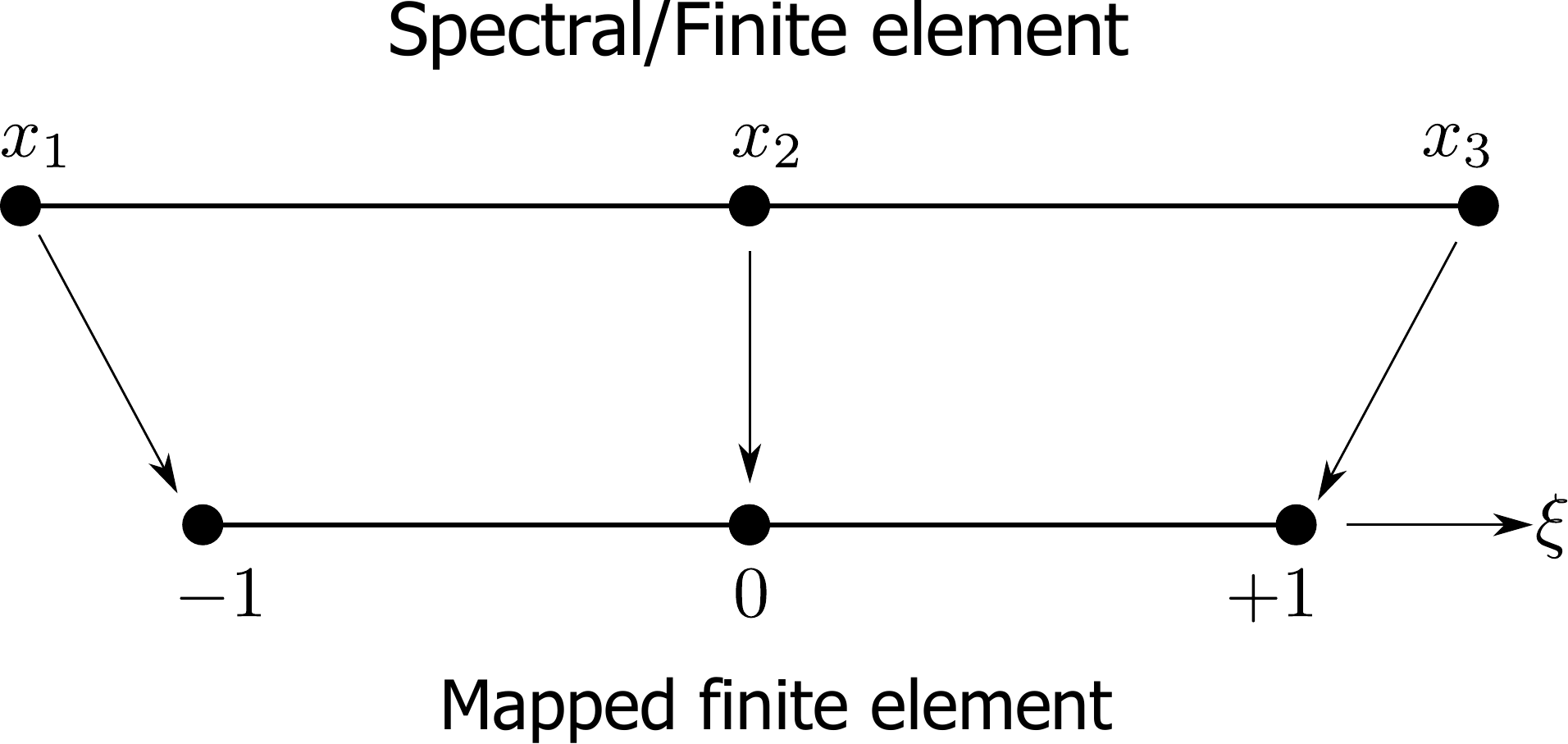}} \hspace{2.0em}
\subfloat[]{\includegraphics[scale=0.28]{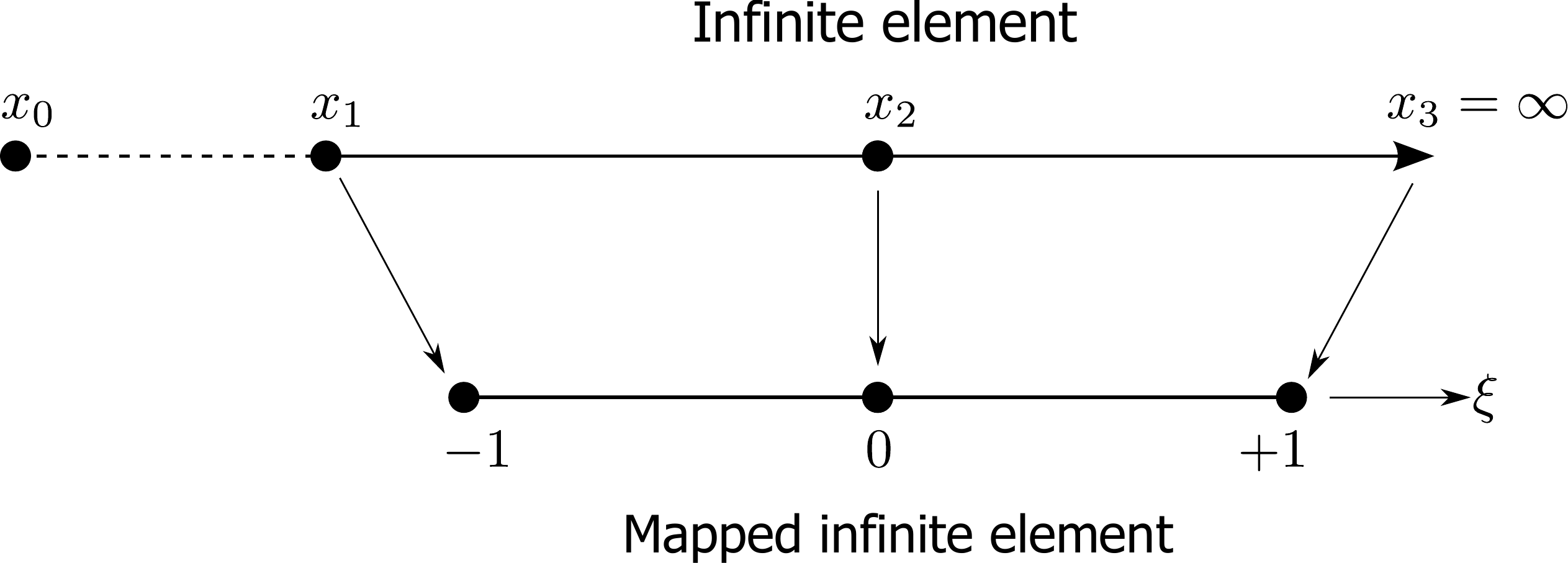}}
\caption{Mapping of a 1D element to the natural coordinates. The element consists of three geometrical nodes. a) Spectral/finite element. b) Infinite element.}
\label{fig:infinite_element1d}
\end{figure}

\begin{figure}[ht]
\centering
\subfloat[]{\includegraphics[scale=0.27]{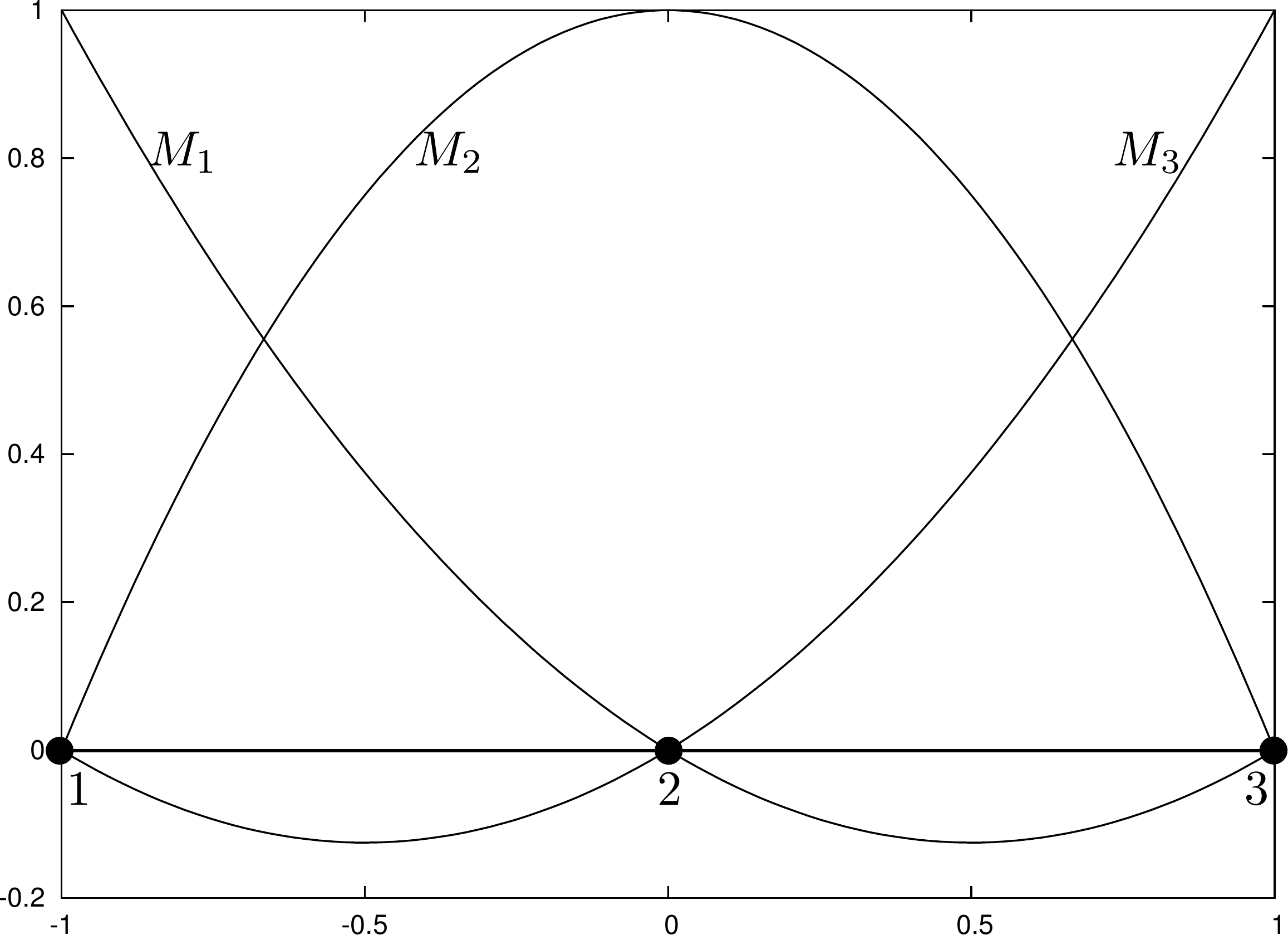}} \hspace{2.0em}
\subfloat[]{\includegraphics[scale=0.27]{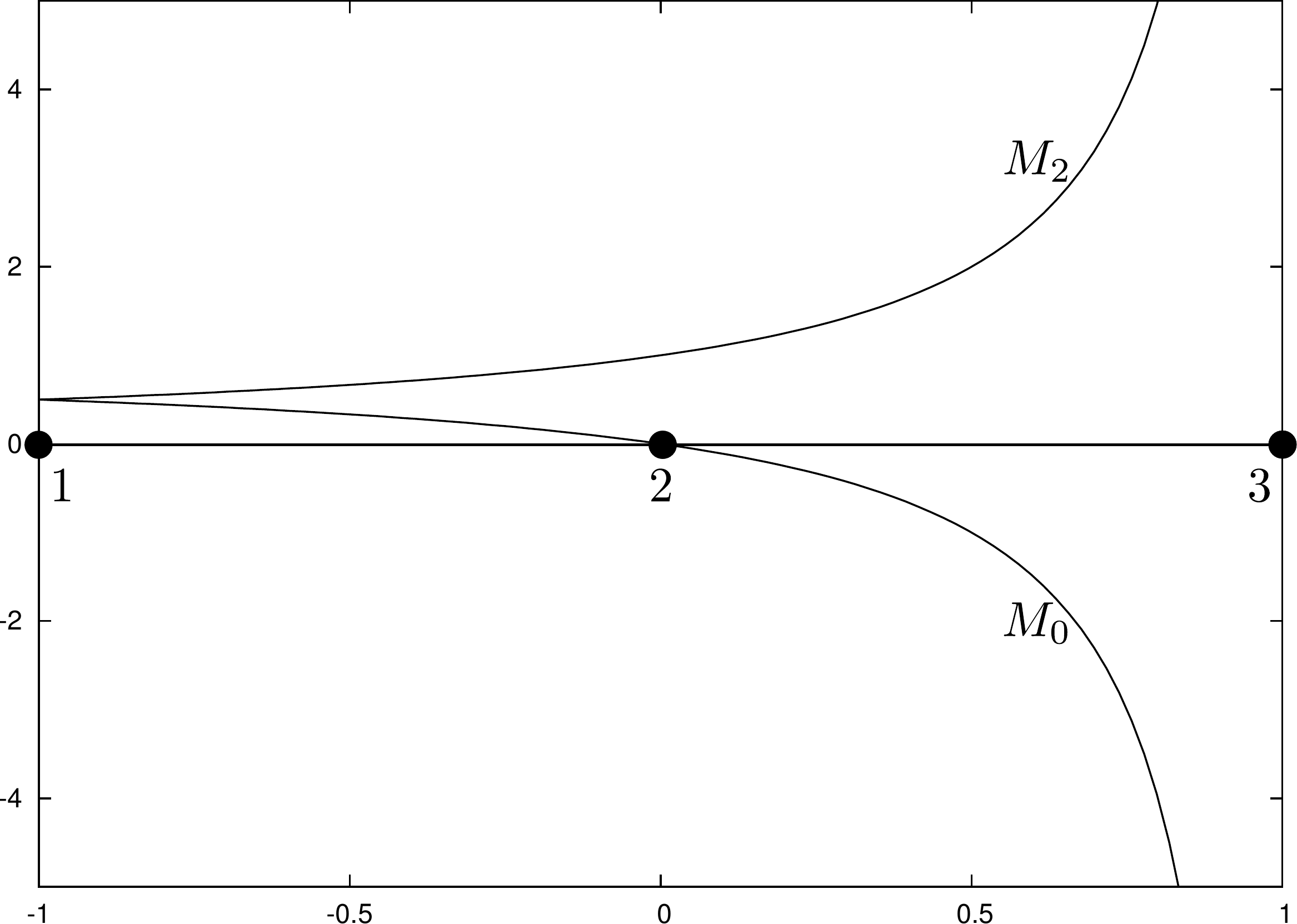}}
\caption{Shape functions for a 1D element with three geometrical nodes. (a) Spectral/finite element. (b) Infinite element.}
\label{fig:shape}
\end{figure}

Inside the domain~$\Omega$, geometrical nodes are used to map an element from the physical domain to the natural domain (Figure~\ref{fig:infinite_element1d}a). 
Outside the domain~$\Omega$, we introduce a single layer of elements in which the gravity field is discretized using the infinite elements (Figure~\ref{fig:inflayer}).
For simplicity and clarity, we illustrate a one-dimensional (1D) mapping.
A point known as the pole, $x_0$, and an intermediate geometrical node, $x_2$, are used to map the element from the physical domain to the natural domain (Figure~\ref{fig:infinite_element1d}b) using the transformation  \cite{zienkiewicz1983,curnier1983}
\begin{equation}
x=M_0(\xi)\,x_0+M_2(\xi)\,x_2
\quad,
\label{eq:infmap1d}
\end{equation}
where the shape functions $M_0(\xi)$ and $M_2(\xi)$ are defined as 
\begin{equation}
\begin{split}
M_0(\xi) &= \frac{\mbox{}-\xi}{1-\xi} \quad , \\
M_2(\xi) &= 1+\frac{\xi}{1-\xi}
\quad.
\end{split}
\label{eq:infshape}
\end{equation}
The shape functions $M_0(\xi)$ and $M_2(\xi)$ satisfy the relation
\begin{equation}
M_0(\xi)+M_2(\xi)=1
\quad.
\label{eq:unity}
\end{equation}
These shape functions map an element in the physical domain, which is extended to infinity, to an element in the natural domain in a following manner (Figure~\ref{fig:infinite_element1d}b):
\begin{equation}
\begin{split}
\xi&=\mbox{}-1\qquad x =x_1 \quad, \\
\xi&=0\qquad\,\,\,\,\, x =x_2 \quad, \\
\xi&=\mbox{}+1\qquad x =x_3=\infty
\quad.
\end{split}
\end{equation}
For infinite elements, shape functions therefore become singular on right-end boundary nodes.
Shape functions for regular spectral and infinite elements are plotted in Figure~\ref{fig:shape}.

To understand the consequence of infinite-element shape functions, we can express the field variable as a piecewise polynomial
\begin{equation}
\Phi(\xi)=a_0+a_1\,\xi+a_2\,\xi^2+a_3\,\xi^3+ \cdots
\quad .
\label{eq:pxi}
\end{equation}
From equations~(\ref{eq:infmap1d}) and (\ref{eq:infshape}) we determine the inverse map
\begin{equation}
\xi=1-\frac{\gamma\,(x_1-x_0)}{r}
\quad,
\end{equation}
where $\gamma=\left(x_2-x_0\right)/\left(x_1-x_0\right)$ is a constant, given $x_0$, $x_1$, and $x_2$,
and $r=x-x_0$.
The value of $\gamma$ governs the location of the pole, $x_0$, and hence the size of the infinite element.

\begin{figure}[ht]
\centering
\subfloat[]{\includegraphics[scale=0.32]{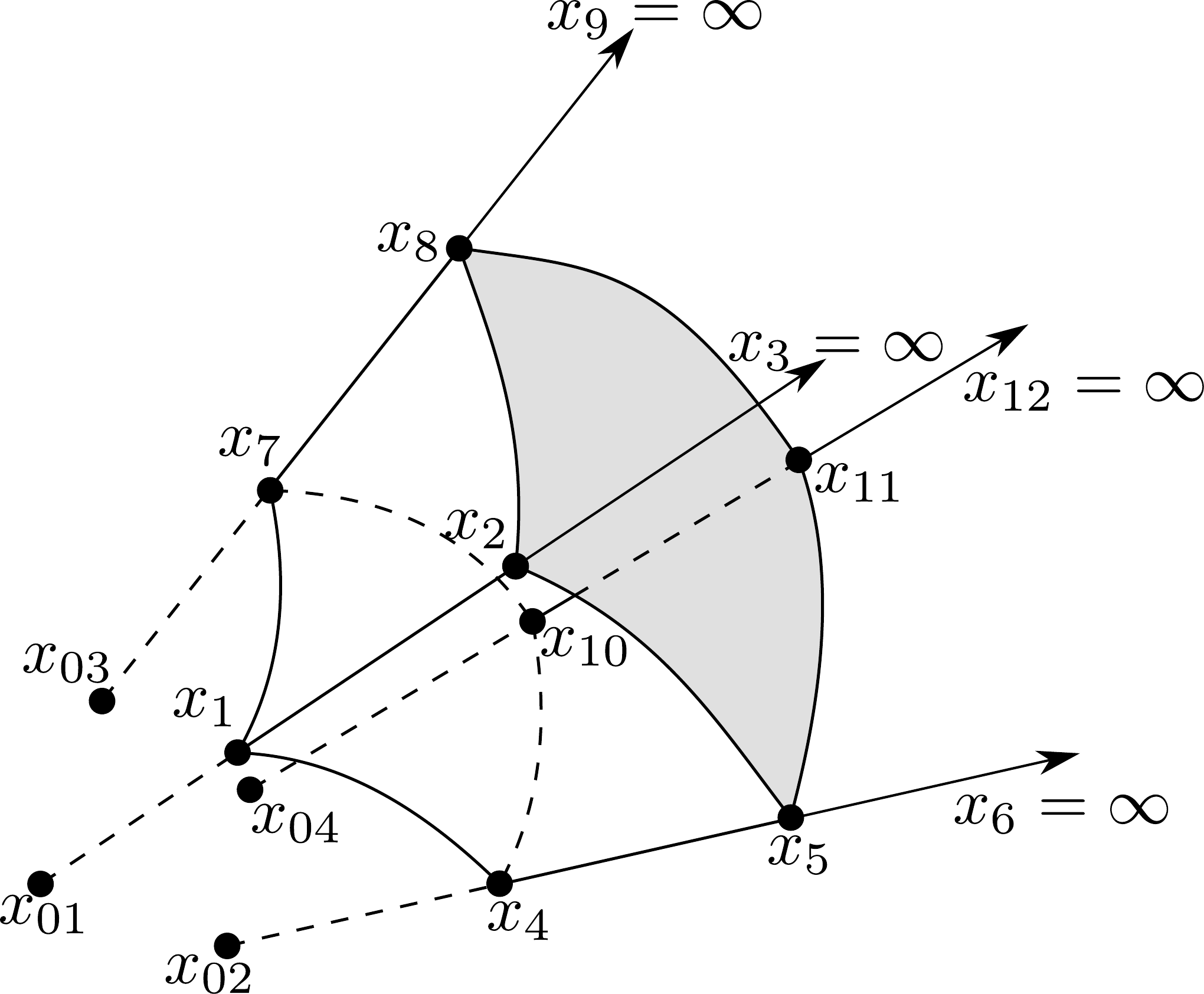}} \hspace{2.0em}
\subfloat[]{\includegraphics[scale=0.32]{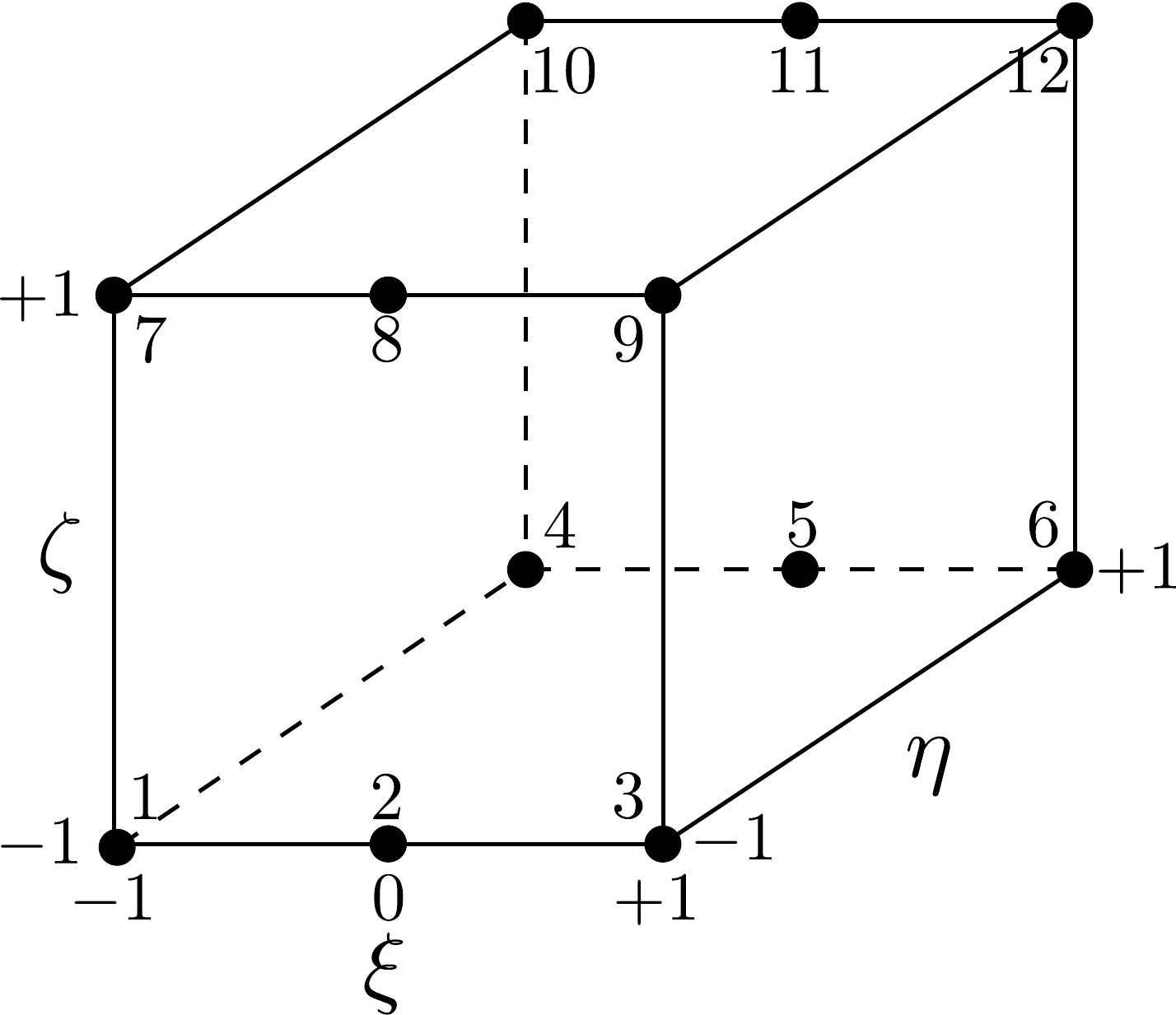}}
\caption{(a) A general 3D infinite element. Points $x_{01}$, $x_{02}$, $x_{03}$, and $x_{04}$ are the poles. (b) 3D infinite element mapped to its natural coordinates.}
\label{fig:infinite_element3d}
\end{figure}

Substituting $\xi$ in equation~(\ref{eq:pxi}), we obtain 
\begin{equation}
\Phi(r)=b_0+\frac{b_1}{r}+\frac{b_2}{r^2}+\frac{b_3}{r^3}+\cdots\quad,
\end{equation}
where, $b_0$, $b_1$, $b_2$, $\cdots$ are constants for given $x_0$, $x_1$, and $x_2$. Hence, the value of $\Phi(r)$ decays to $b_0$ as $r$ tends to $\infty$.
For our problem, since gravity decays to zero at $\infty$, the value of $b_0$ is zero.
One or more poles may be necessary depending on the physics and the model.
All poles have to be located opposite to the decay direction and outside the infinite element.
The accuracy of the infinite-element approximation may be increased by increasing the order of the interpolation functions, but the shape functions remain the same. To maintain a positive Jacobian across all the integration points, infinite elements must have either parallel or diverging faces toward infinity. Shape functions in two or three dimensions may be obtained using tensor products similar to equation~(\ref{eq:lag}).
A typical mapping of a 3D infinite element in a general physical domain to an element in the natural domain is shown in Figure~\ref{fig:infinite_element3d}.

Alternatively, the shape functions for the infinite element can be defined using Node~1 and Node~2~\cite{kumar1985,marques1984}. Similarly, the shape functions for the particular decay functions, such as exponential or logarithmic decay, can also be derived~\cite{abdelfattah2000}.

\subsection{Numerical integration}

\begin{figure}[ht]
\centering
\includegraphics[scale=0.5]{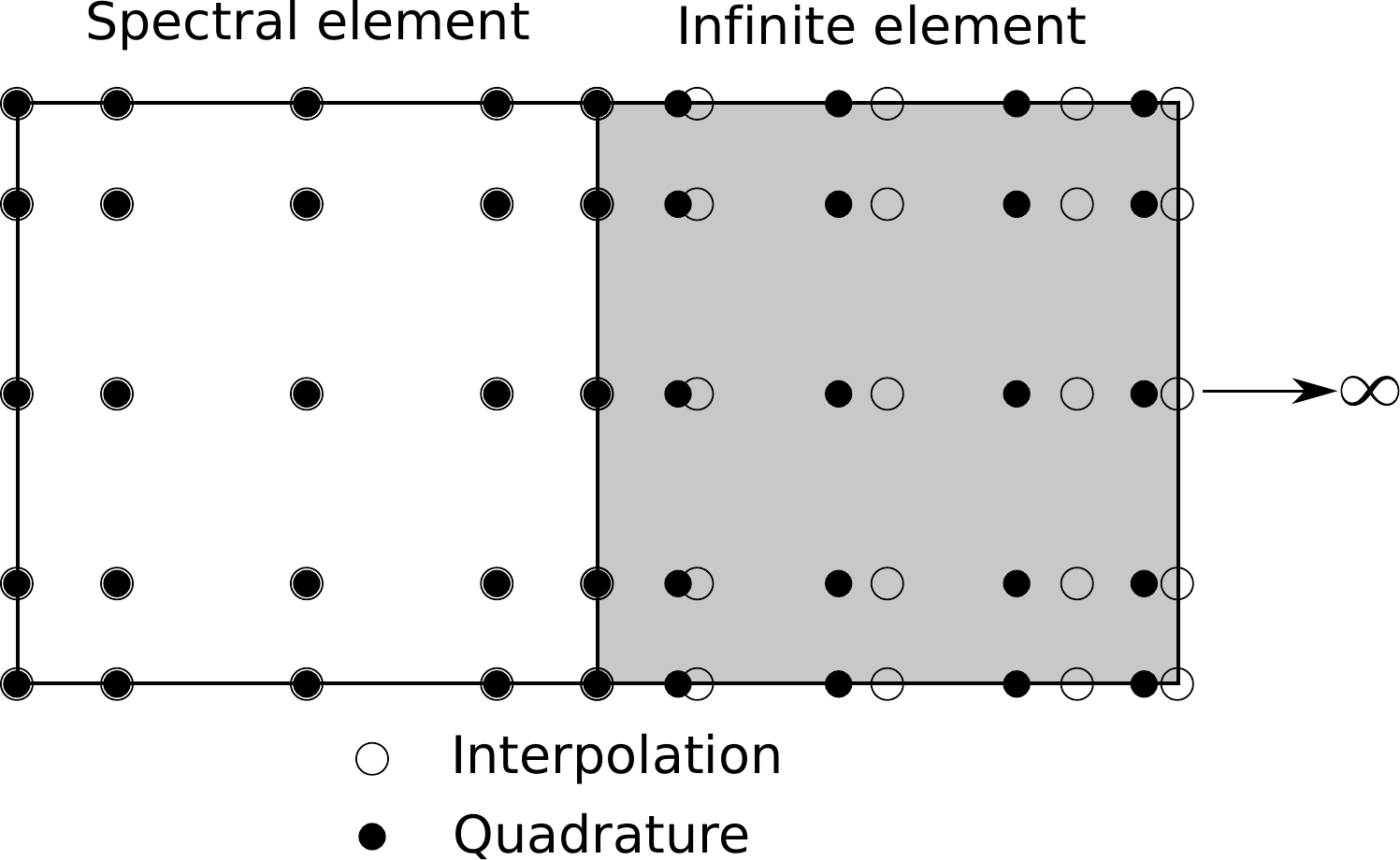}
\caption{Coupling between a spectral and an infinite element in 2D. Both elements use identical interpolation nodes. Spectral elements use Gauss-Lobatto-Legendre (GLL) quadrature points, whereas infinite elements use Gauss-Radau (GR) quadrature points in the infinite direction and GLL quadrature in the two remaining directions.
GLL and GR quadrature points coincide on a spectral-infinite element boundary.}
\label{fig:shape12}
\end{figure}

In a spectral element we use Gauss-Legendre-Lobatto (GLL) quadrature:
\begin{equation}
\int_{-1}^1f(\xi)\,\rmd\xi=w_1\,f(-1)+\sum_{\alpha=2}^{n-1}w_{\alpha}\,f(\xi_\alpha)+w_{n}\,f(1)
\quad,
\end{equation}
where~$f$ is a general function, $w_\alpha$ are the GLL weights of integration, and $\xi_\alpha$ are the quadrature points.
GLL quadrature is exact for polynomials of order $2\,n-3$ or less.
Since the quadrature includes the end points of the interval, shape functions for infinite elements are singular at the infinite boundary of the element.
Consequently, GLL quadrature cannot be used in infinite element.

One option is to use the Gauss-Legendre quadrature, which is given by
\begin{equation}
\int_{-1}^1f(\xi)\,\rmd\xi=\sum_{\alpha=1}^{n}w_{\alpha}\,f(\xi_\alpha)
\quad.
\end{equation}
Gauss-Lengendre quadrature is exact for polynomial of order $2\,n-1$ or less.
Since the quadrature does not include the end points of the interval, infinite-element shape functions can be computed at all quadrature points.
Another option is to use the Gauss-Radau quadrature, which is given by 
\begin{equation}
\int_{-1}^1f(\xi)\,\rmd\xi=w_1\,f(-1)+\sum_{\alpha=2}^{n}w_{\alpha}\,f(\xi_\alpha)
\quad.
\end{equation}
Gauss-Radau quadrature is exact for polynomials of order $2\,n-2$ or less,
and includes only the near end points of the interval.
Therefore, infinite element shape functions can be safely computed at all quadrature points.
By combining Gauss-Radau quadrature in infinite elements with GLL quadrature in spectral elements, quadrature points on an spectral-infinite element interface coincide (Figure~\ref{fig:shape12}). This coincidence naturally couples spectral and infinite elements. Therefore, we use the Gauss-Radau quadrature in infinite elements.

\section{Parallelization}

We use non-overlapping domain decomposition for parallelization, in which each partition contains a unique set of elements,
and nodes are only shared on interfaces.
Since pre- and post-processing of infinite elements is similar to spectral elements,
infinite elements do not pose any difficulty for parallelization.
Depending on the geometry of the domain decomposition, the pole of an infinite element may lie in other partitions, which does not pose any problem for parallelization.
We implemented parallel iterative Krylov solvers using PETSc, a portable and extensible toolkit for scientific computation~\cite{petsc2015}.

\section{Numerical examples}

\subsection{Homogeneous sphere}

\begin{figure}[htbp]
\centering
\includegraphics[scale=0.17]{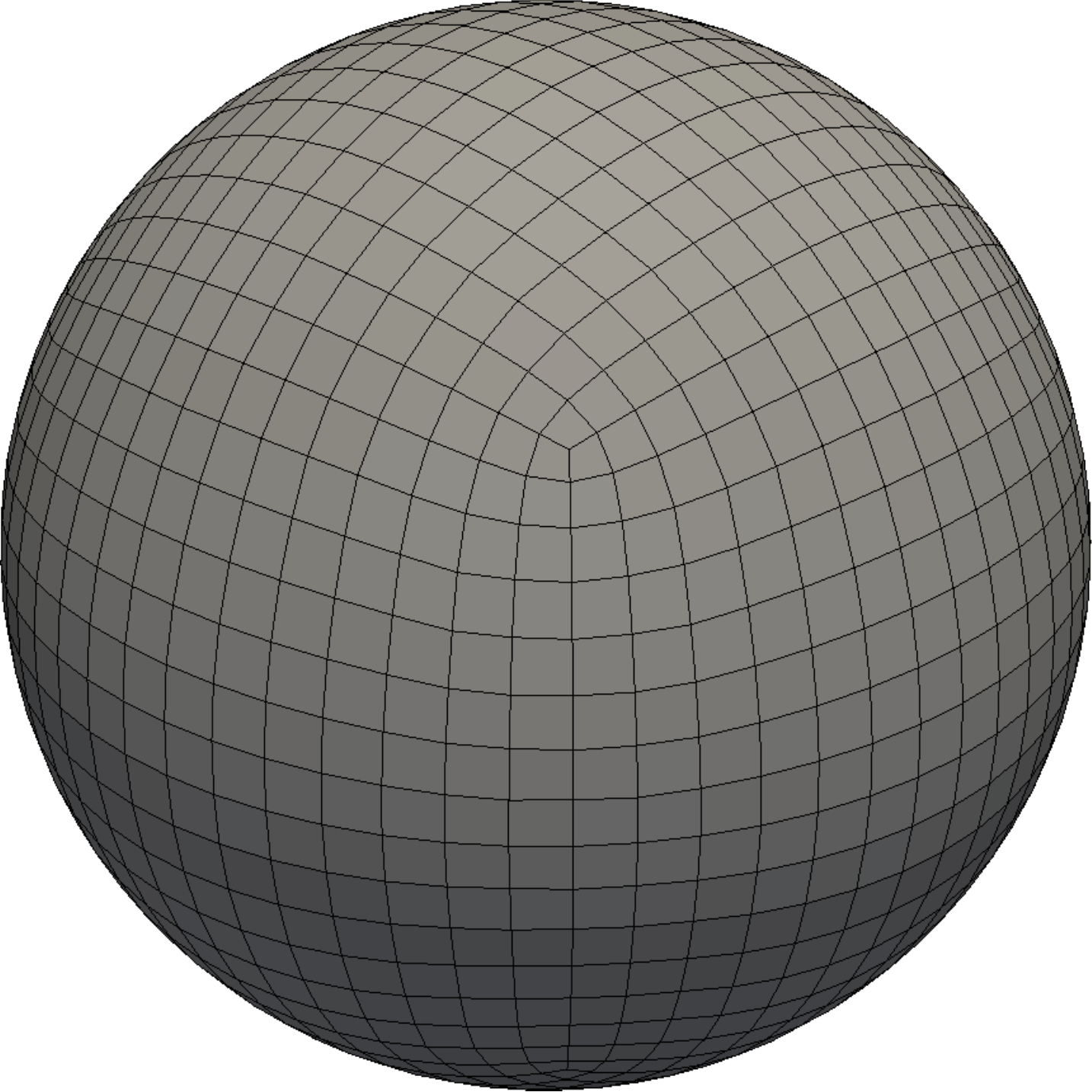}
\caption{Mesh for the homogeneous sphere, which has a radius of~$1000$~m and a mass density of~$1.92$~kg/m\textsuperscript{3}.}
\label{fig:sphere}
\end{figure}

\begin{figure}[htbp]
\centering
\subfloat[]{\includegraphics[scale=0.30]{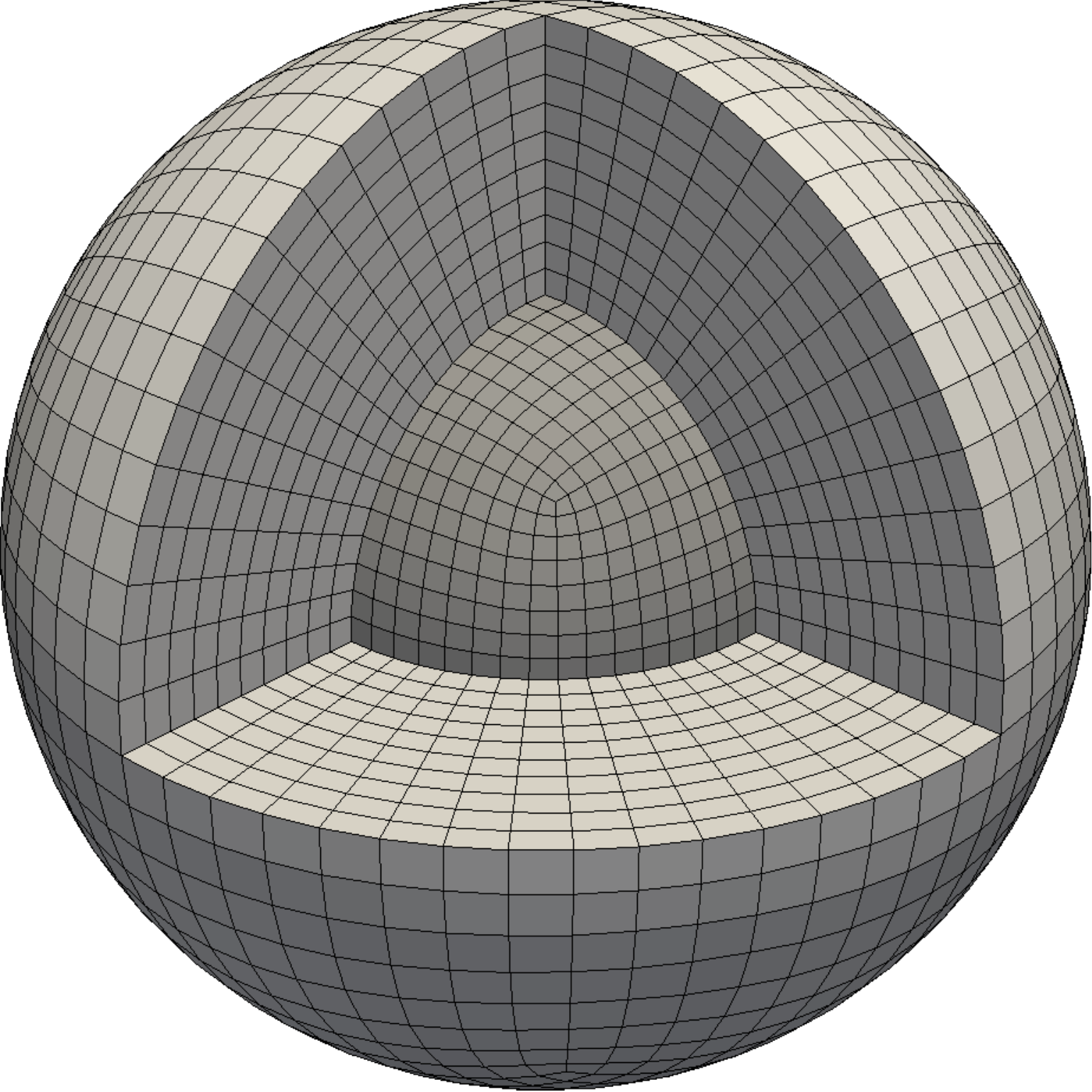}}\hspace{2.0em}
\subfloat[]{\includegraphics[scale=0.60]{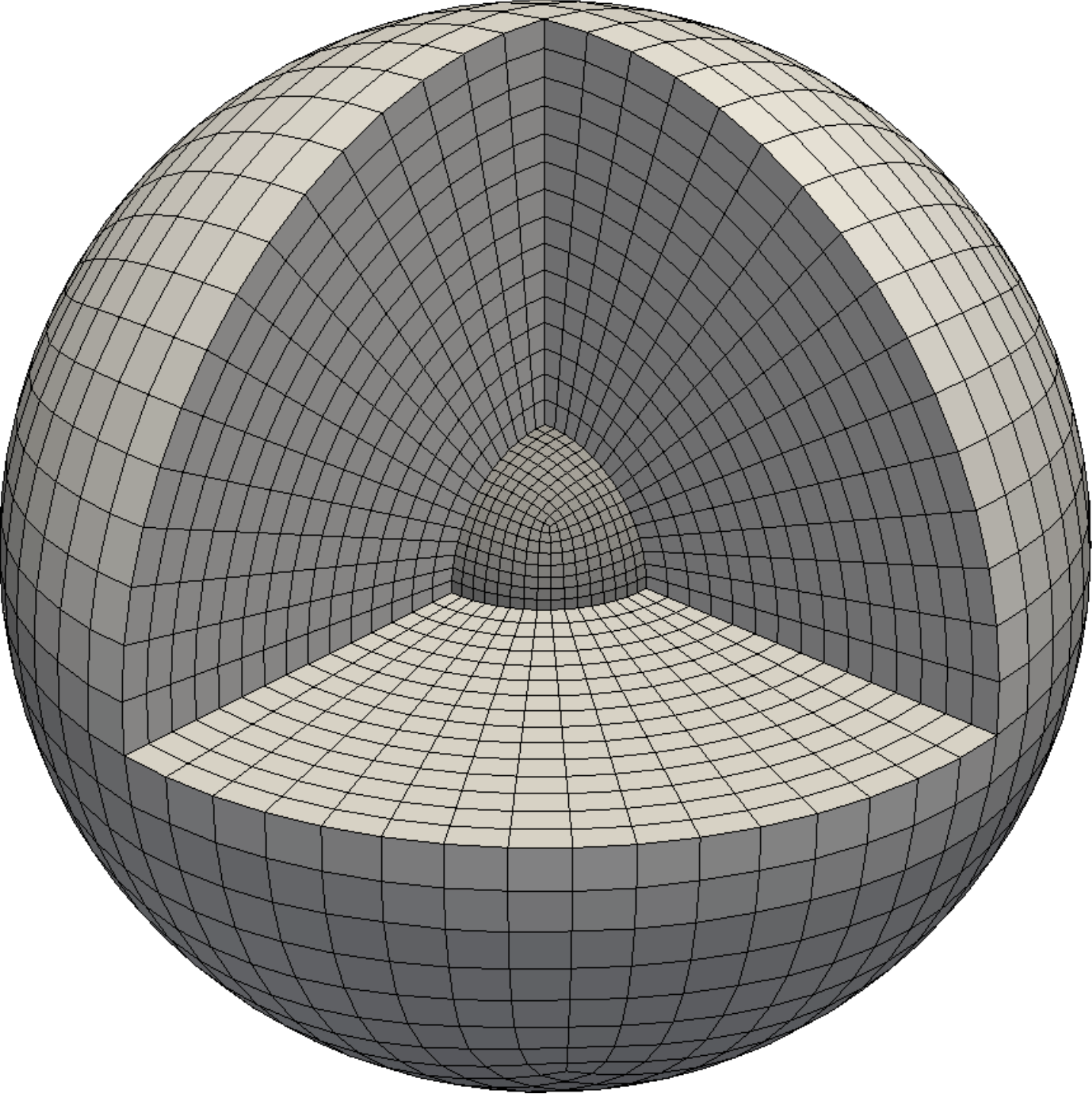}}
\caption{Mesh for the homogeneous sphere with an outer layer of thickness a) $1000$ and b) $3000$. The sphere has a mass density of $1.92$~kg/m\textsuperscript{3},
and the outer layers have zero mass density.}
\label{fig:sphere_ext2}
\end{figure}

\begin{figure}[htbp]
\centering
\includegraphics[scale=0.17]{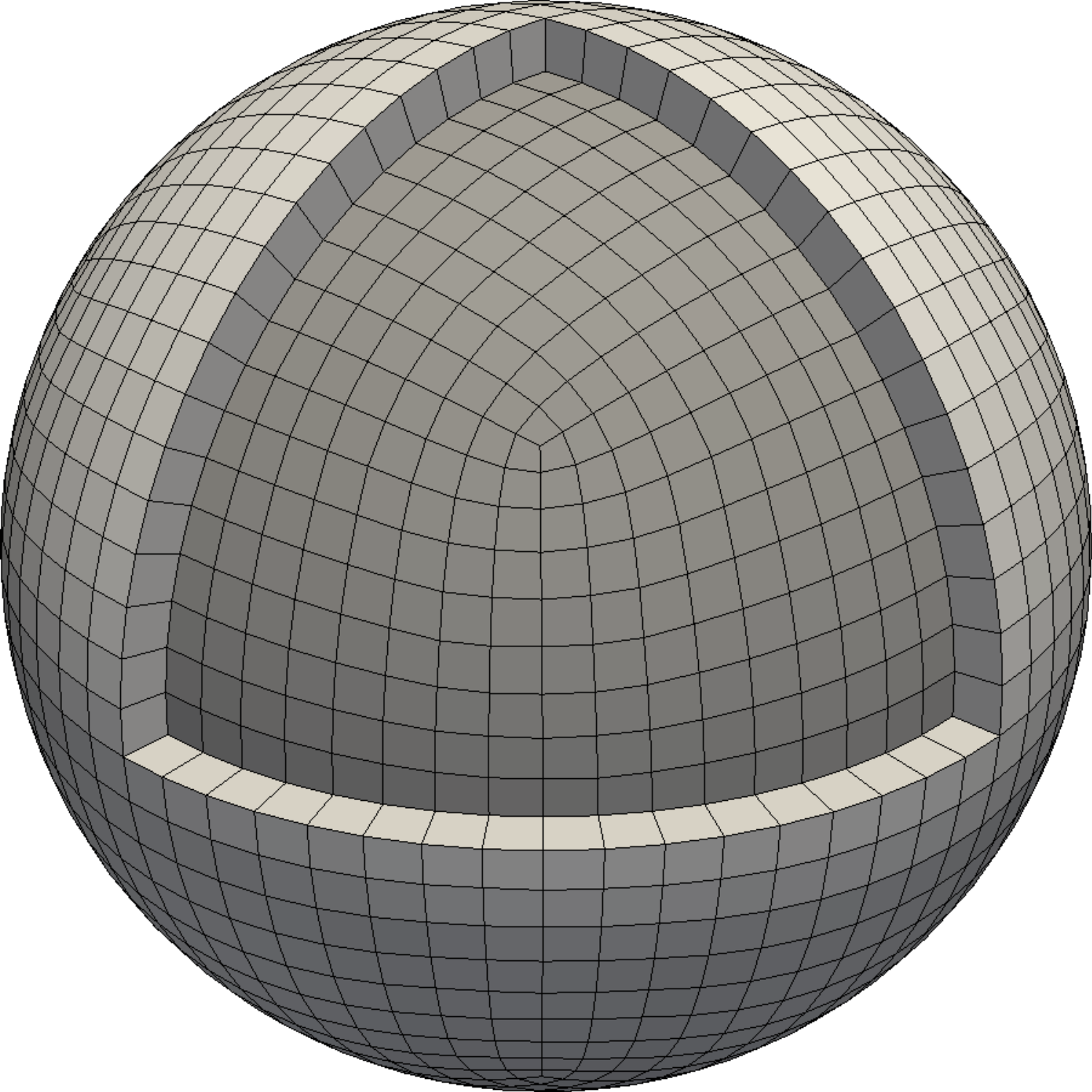}
\caption{Mesh for the homogeneous sphere with an infinite-element layer. The sphere has a mass density of $1.92$~kg/m\textsuperscript{3} and the infinite-element layer has zero mass density.}
\label{fig:sphere_inf}
\end{figure}

In the first example, we consider a homogeneous sphere with a radius $1000$~m and a mass density of $1.92$~kg/m\textsuperscript{3}.
The analytical solution for the gravitational potential of the homogeneous sphere is given by
\begin{equation}
\Phi(r)=
\begin{cases}
\frac{2}{3}\pi\,G\,\rho\left(3\,r_0^2-r^2\right) & \text{for }r\le r_0 \quad ,\\
\frac{4}{3}\pi\,G\,\rho\,\frac{r_0^3}{r} & \text{for }r> r_0 \quad,
\end{cases}
\label{eq:analytical}
\end{equation}
where $r_0$ is the radius of the sphere.
The volume of the sphere is meshed using hexhedral elements, as shown in Figure~\ref{fig:sphere}.
We use an element size of 10~m, and the outer layer is meshed with a slightly coarser mesh. For quasistatic problems, three GLL points may be sufficient for accurate simulation~\cite{gharti2012a}. Therefore, we use three GLL points in this article. We solve the unbounded Poisson's equation considering four different cases as follows:
\renewcommand\labelenumi{\textbf{Case \theenumi.}}
\begin{enumerate}
\item Sphere of radius 1000~m (Figure~\ref{fig:sphere}).
\item Sphere of radius 1000~m and an outer layer of thickness 1000~m (Figure~\ref{fig:sphere_ext2}a).
\item Sphere of radius 1000~m and an outer layer of thickness 3000~m (Figure~\ref{fig:sphere_ext2}b).
\item Sphere of radius 1000~m and a correct Dirichlet boundary condition on the surface (Figure~\ref{fig:sphere}).
\item Sphere of radius 1000~m and a single infinite element layer (Figure~\ref{fig:sphere_inf}).
\end{enumerate}

\begin{figure}[htbp]
\centering
\subfloat[]{\includegraphics[scale=0.37]{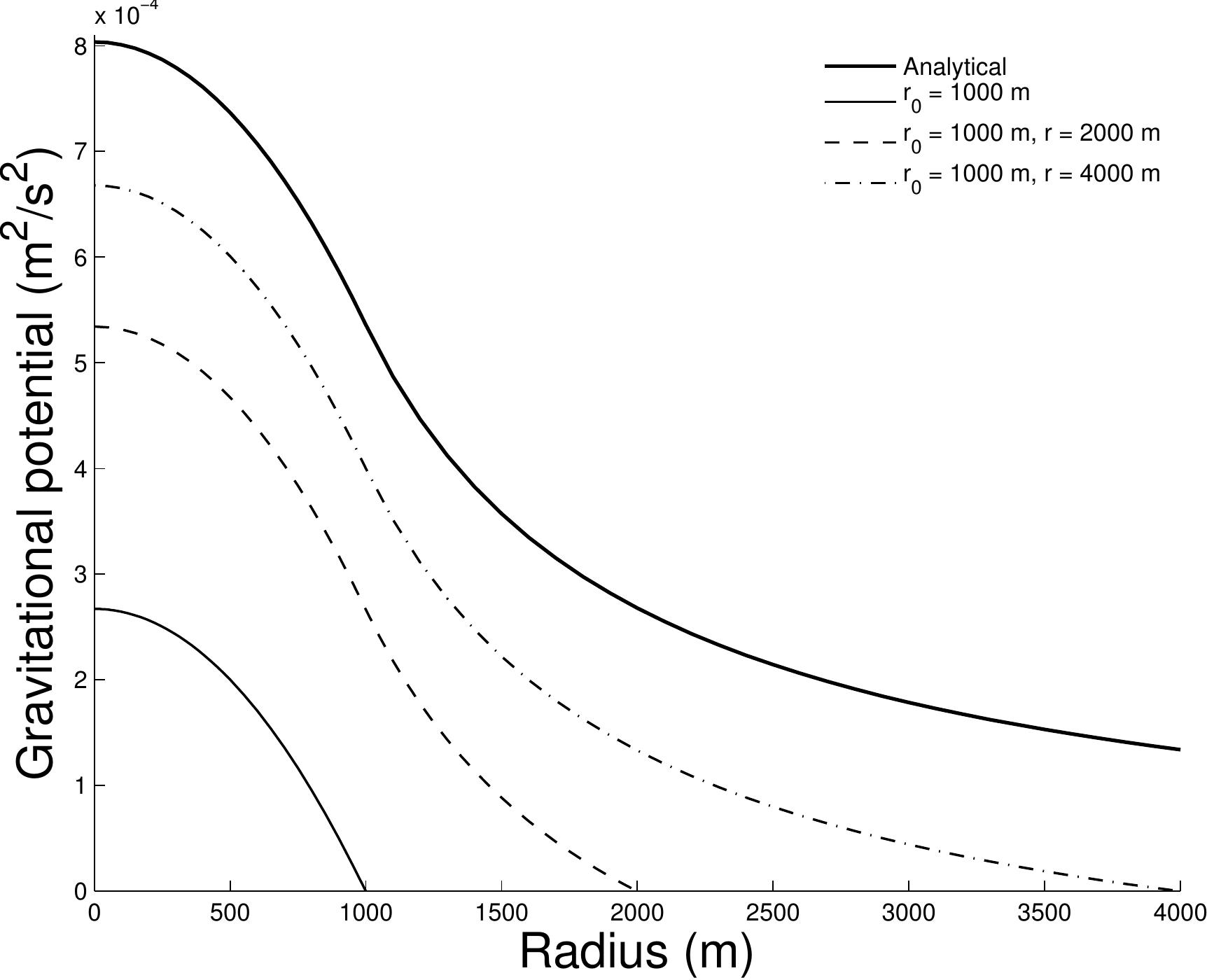}}\hspace{2.0em}
\subfloat[]{\includegraphics[scale=0.37]{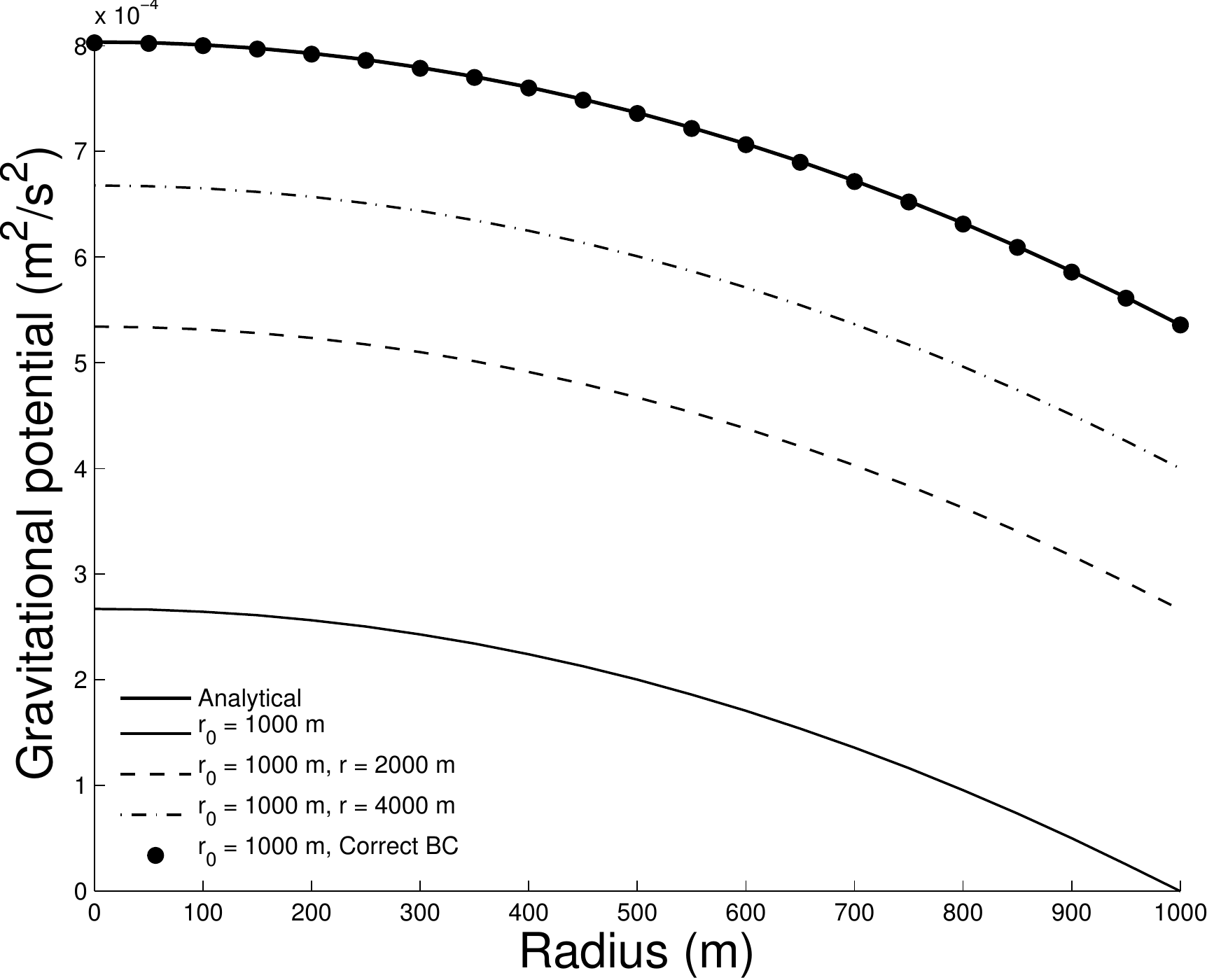}}
\caption{(a) Radial profile of the gravitational potential. (b) Same as (a) but compared with the result based on the correct boundary conditions.}
\label{fig:phi_finite}
\end{figure}

Figure~\ref{fig:phi_finite}a shows the results for the first three cases using the regular spectral-element method with the boundary condition $\Phi=0$ on the outer surface. The results are compared with the analytical solution~( Equation \ref{eq:analytical}).
As expected, if we assume $\Phi=0$ on the outer surface of the original domain, the result is far from the analytical solution.
As we extend the model domain, the numerical solution approaches the analytical solution, but the model has to be extended significantly.
Figure~\ref{fig:phi_finite}b shows the result for the original model assuming that the correct boundary conditions are known on the surface. In this case, the result perfectly matches the analytical solution. Hence, if accurate boundary conditions are known on the surface of the model, the regular spectral-element method accurately computes the gravity field.

\begin{figure}[htbp]
\centering
\subfloat[]{\includegraphics[scale=0.37]{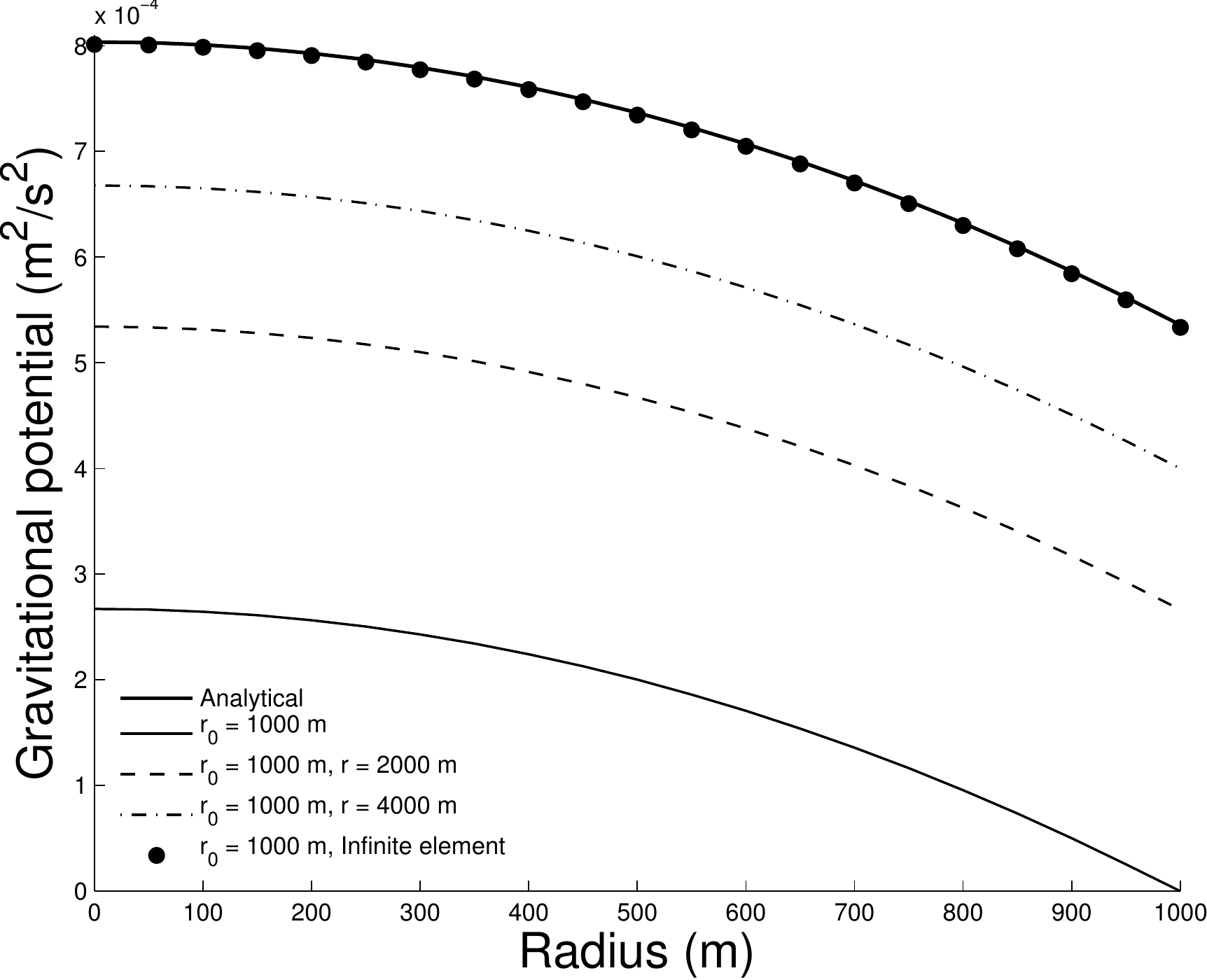}}\hspace{2.0em}
\subfloat[]{\includegraphics[scale=0.37]{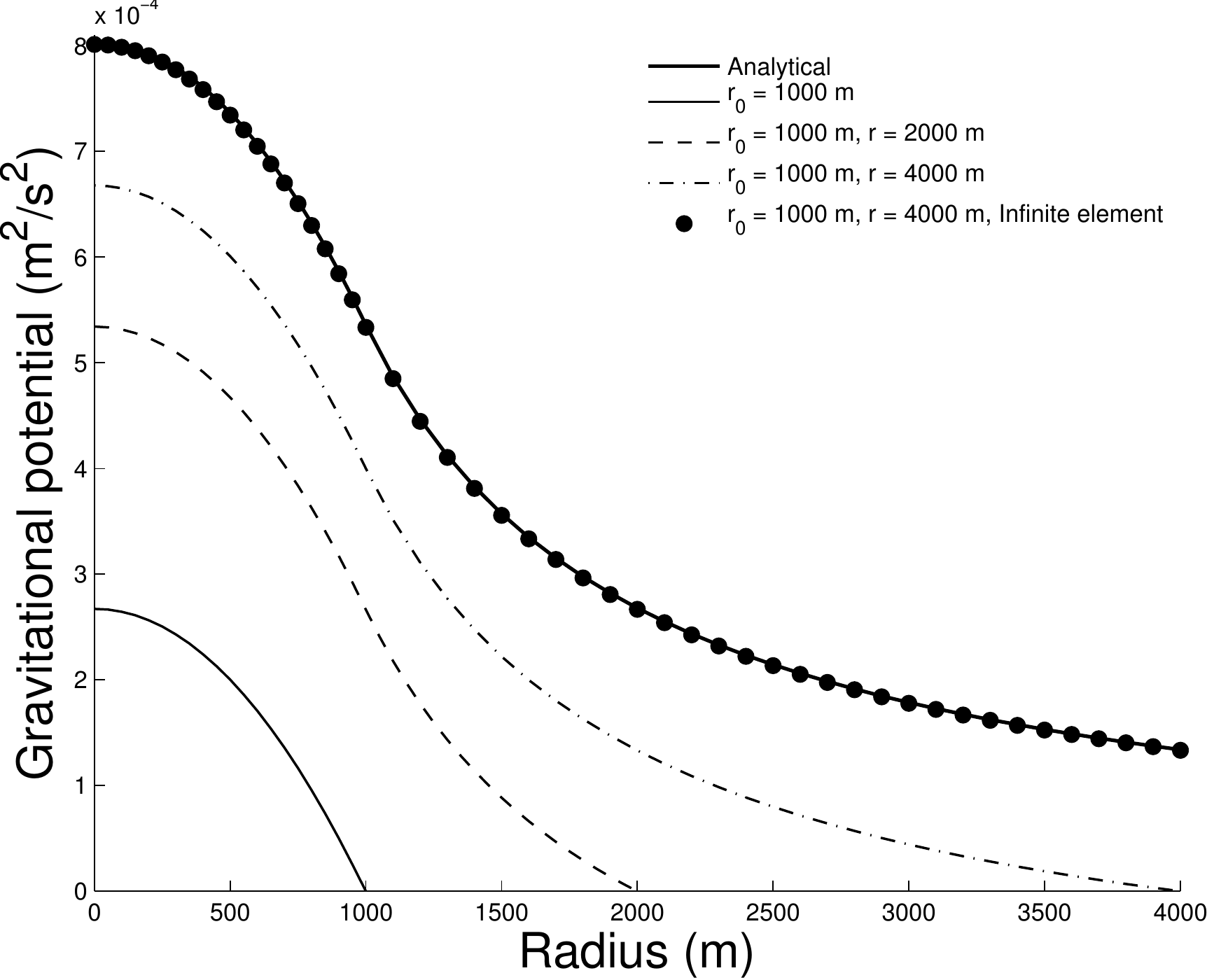}}
\caption{Radial profile of the gravitational potential compared with (a) an infinite-element layer on top of the original model and (b) an infinite-element layer on top of the extended model $r=4000$~m.}
\label{fig:phi_infinite}
\end{figure}

Finally, Figure~\ref{fig:phi_infinite}a shows the result for an infinite element layer added on top of the original model,
which perfectly matches the analytical solution.
To test the accuracy of the infinite-element method in outer space, we added an infinite element layer on top of the extended model with an outer layer thickness of 3000~m. In this case also the result matches very well with the analytical solution both inside and outside the domain, as shown in Figure~\ref{fig:phi_infinite}b.

\subsection{Preliminary Reference Earth Model}

\begin{figure}[htbp]
\centering
\subfloat[]{\includegraphics[scale=0.36]{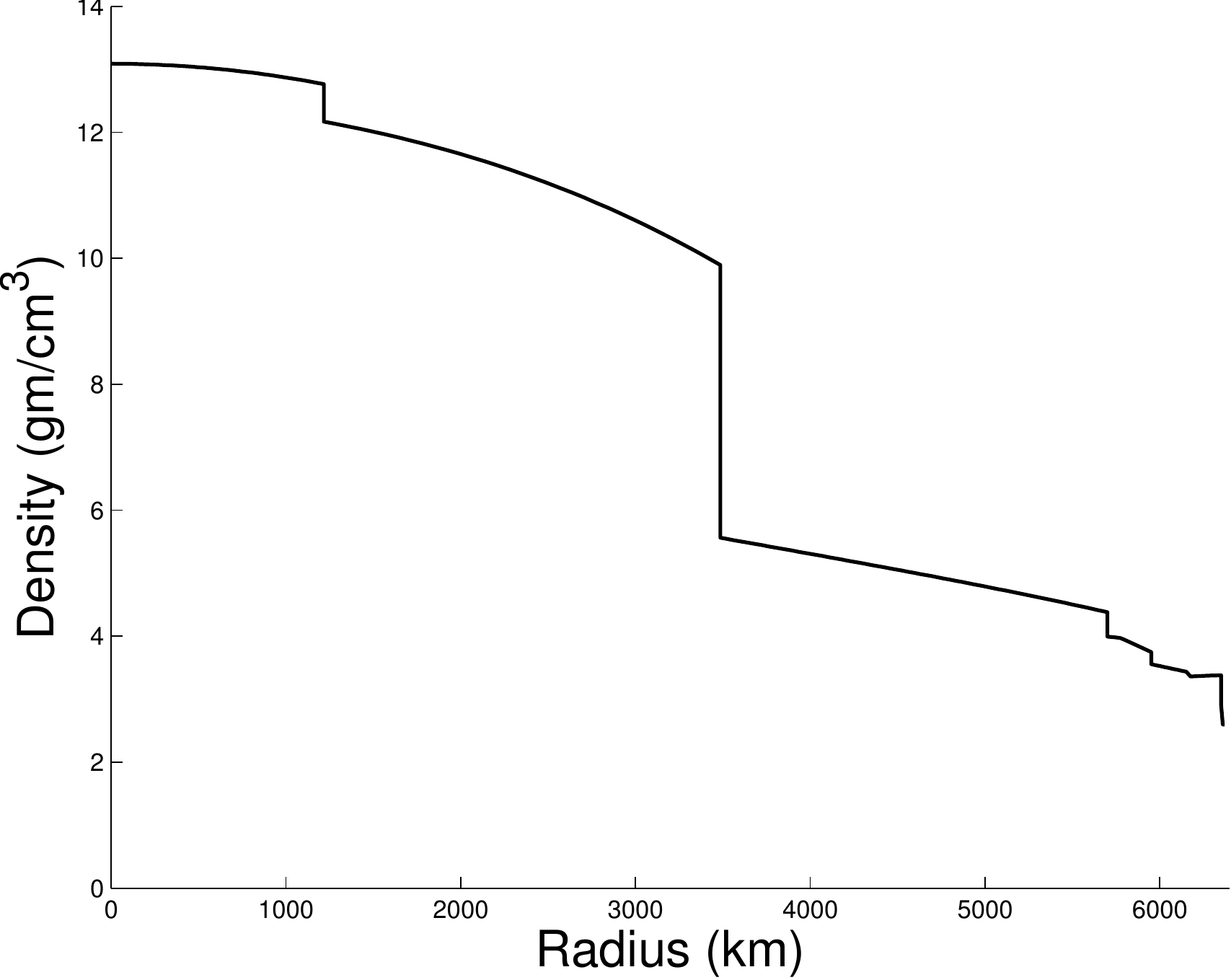}}\hspace{2.0em}
\subfloat[]{\includegraphics[scale=0.36]{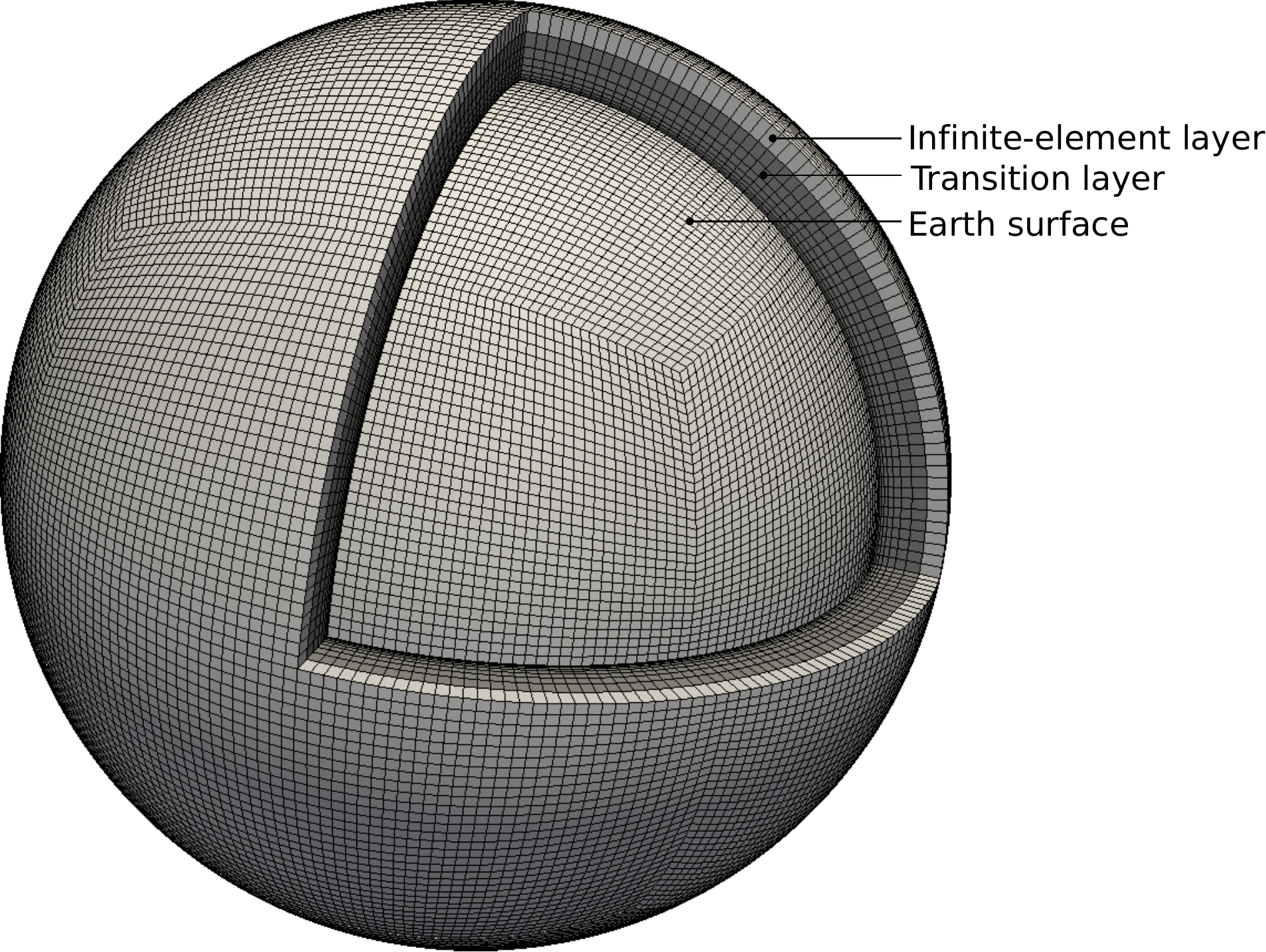}}
\caption{(a) Mass density profile for the Preliminary Reference Earth Model (PREM) \cite{dziewonski1981}. (b) Mesh of the Earth model with added transition layer and infinite-element layer.}
\label{fig:prem}
\end{figure}

\begin{figure}[htbp]
\centering
\subfloat[]{\includegraphics[scale=0.36]{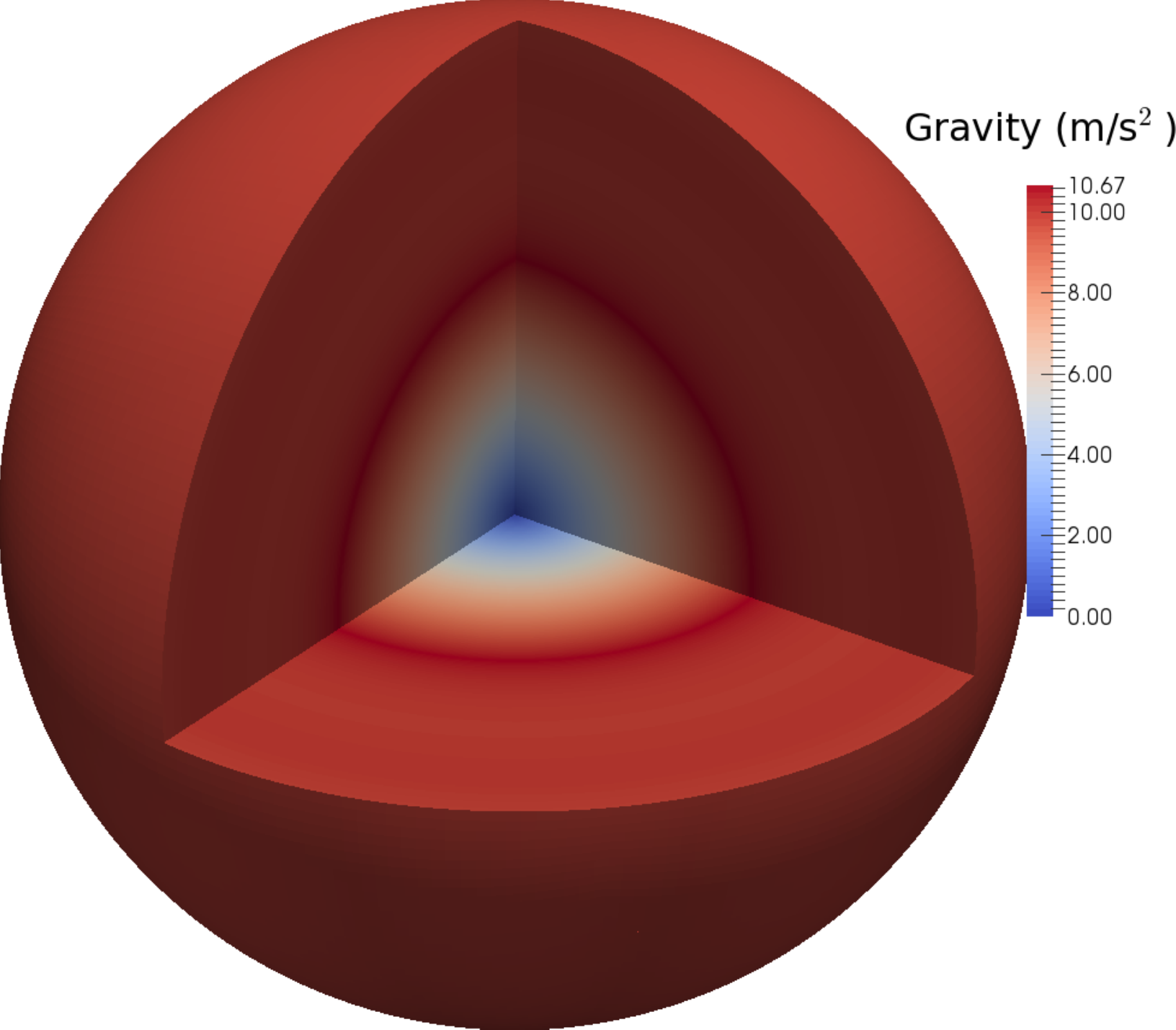}}\hspace{2.0em}
\subfloat[]{\includegraphics[scale=0.36]{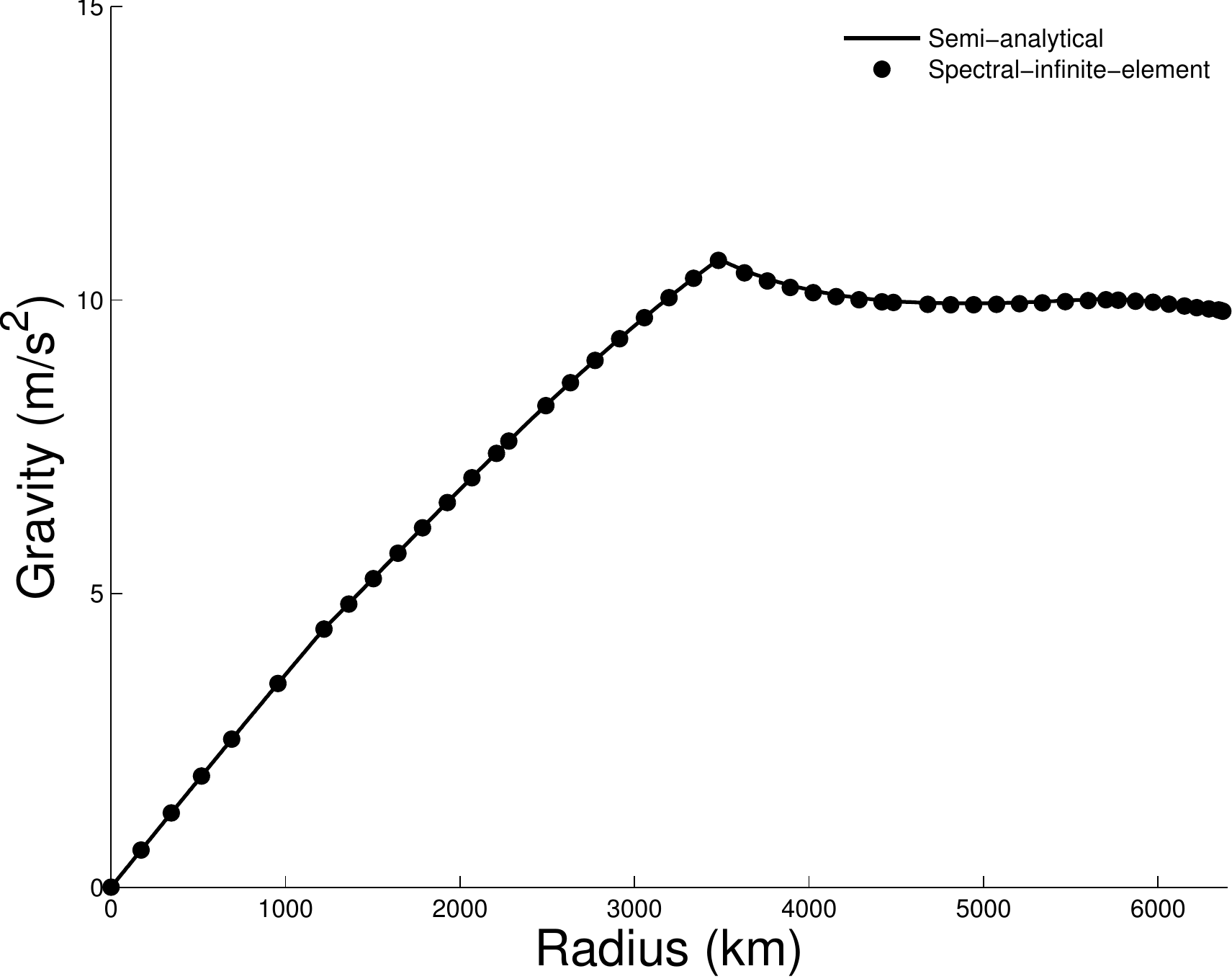}}
\caption{(a) Gravitational acceleration computed for PREM (Figure~\ref{fig:prem}a). (b) Radial profile of the gravitational acceleration.}
\label{fig:prem_result}
\end{figure}

As a more realistic test,
we compute the gravitational acceleration in the Preliminary Reference Earth Model (PREM)~\cite{dziewonski1981}, as show in Figure~\ref{fig:prem}a.
We again mesh the Earth model using hexahedral elements. All internal discontinuities are honored by the mesh.
To avoid inaccuracies on the Earth's surface, we added a transition layer between the Earth and the infinite element layer (Figure~\ref{fig:prem}b). The mesh is decomposed based on the nonoverlapping domian decomposition for parallel processing. Figure~\ref{fig:prem_result} shows the results computed using the parallel infinite-spectral-element method, which compare favorably with the semi-analytical solution.
Gravity is zero at Earth's center, and increases almost linearly towards the core-mantle boundary, where it reaches its maximum.
Gravity remains approximately constant up to the Earth's surface.
Even though we use only three GLL points per element and a relatively courser mesh, the results are very accurate.

\section{Conclusions}

We have successfully implemented an infinite-spectral-element method to solve the unbounded Poisson/Laplace equation.
The results are very accurate and the method is highly efficient. The result of the dynammic simulation shows that the method is stable. The main advantage of this numerical discretization of the unbounded Poisson equation is that it can be naturally coupled with the the laws of
continuum mechanics that govern geostatic and geodynamic deformations.
This allows us to address several geophysical problems, such as long period seismic wave propagation, glacial-isostatic adjustment, postseismic relaxation, and tidal loading;
and future work will focus on these important problems.

In this article, we used only three GLL points per element, which gives accurate results. However, it has been shown that five GLL points are usually necessary for dynamic problems, for example, in simulations of seismic wave propagation~\cite{komatitsch2002a}.
In such cases, we have two choices: 1) use three GLL points for the gravity field and five GLL points for the displacement field, thus requiring a mixed formulation, or 2) use the same number of GLL points for both the gravity and displacement fields.
We have seen that an infinite-element discretization provides implicit coupling with spectral elements, which avoids iterative procedures.
Since there is only a single layer of infinite elements, the additional computational cost is insignificant.

\section{Acknowledgments}

We thank Stefano Zampini for his help on implementing PETSc library. Parallel programs were run on computers provided by the Princeton Institute for Computational Science and
Engineering (PICSciE). 3D data were visualized using the open-source parallel visualization software
ParaView/VTK (www.paraview.org). This research was partially supported by NSF grant 1112906. 

\bibliographystyle{plain}

\end{document}